\documentclass[prl,reprint,onecolumn,aps,amssymb,graphicx,floatfix,amsmath,superscriptaddress,showpacs,longbibliography]{revtex4-1}
\usepackage{bm}
\usepackage{float}
\usepackage{graphicx}
\usepackage{color}
\usepackage{amsfonts}
\usepackage{amsmath}
\usepackage{textcomp}
\usepackage{subcaption}
\usepackage{hyperref}
\usepackage{comment}
\hypersetup{colorlinks=true,linkcolor=blue,anchorcolor=blue,citecolor=blue,filecolor=blue,urlcolor=blue,bookmarksnumbered=true,pdfview=FitB}
\newcommand{\w}{\omega}
\newcommand{\e}{{c(s)}}

\newcommand{\beq}{\begin{equation}}
\newcommand{\eeq}{\end{equation}}
\newcommand{\bea}{\begin{eqnarray}}
\newcommand{\eea}{\end{eqnarray}}

\newcommand{\nn}{\nonumber}

\newcommand{\q}{\mathbf{q}}

\newcommand{\re}{\text{Re}}
\newcommand{\im}{\text{Im}}
\newcommand{\crs}{{\text{cross}}}

\newcommand{\cc}{{\mbox{c.c.}}}
\newcommand{\zs}{{\text{zs}}}
\newcommand{\hzs}{\text{h}}
\newcommand{\mzs}{\text{m}}
\newcommand{\qp}{{\text{qp}}}
\newcommand{\pole}{{\text{pole}}}
\newcommand{\bc}{{\text{branch}}}

\begin{document}

\title{Hidden and mirage collective modes in two dimensional Fermi liquids}
\author{Avraham Klein}
\affiliation{School of Physics and Astronomy, University of Minnesota, Minneapolis, MN 55455}
\author{Dmitrii L. Maslov}
\affiliation{Department of Physics, University of Florida, P.O. Box 118440, Gainesville, Florida 32611-8440, USA}
\author{Andrey V. Chubukov}
\affiliation{School of Physics and Astronomy, University of Minnesota, Minneapolis, MN 55455}
\date{\today}

\begin{abstract}
  We show that a two-dimensional (2D) isotropic Fermi liquid harbors two new types of  collective modes,
  driven by quantum fluctuations,
  in addition to conventional zero sound: ``hidden'' and ``mirage'' modes. The hidden modes occur for relatively weak attractive interaction both in the charge and spin channels with any angular momentum $l$. Instead of being conventional damped resonances within the particle-hole continuum, the hidden modes propagate at velocities larger than the Fermi velocity and have infinitesimally small damping in the clean limit, but are invisible to spectroscopic probes.
  The mirage modes are also propagating modes outside the particle-hole continuum that occur for sufficiently strong repulsion interaction in channels with $l\geq 1$.
  They do give rise to peaks in spectroscopic probes, but are not true poles of the dynamical susceptibility.
  We argue that both hidden and mirage modes occur due to a non-trivial topological structure of the Riemann surface, defined by the dynamical susceptibility. The hidden modes reside below a branch cut that glues two sheets of the Riemann surface, while the mirage modes reside on an unphysical sheet of the Riemann surface. We show that both types of modes give rise to distinct features in time dynamics of a 2D Fermi liquid that can be measured in pump-probe experiments.
  \end{abstract}

\maketitle
{\it \bf {Introduction.~~~}}
Zero-sound (ZS) is a collective excitation of a Fermi liquid (FL) associated with a deformation of the Fermi surface (FS)~\cite{Abrikosov1975,Landau1980,Baym1991,Nozieres1999}. The dispersion of the ZS mode $\omega = v_\zs q$ encodes important information about the strength of correlations, as was demonstrated in classical experiments on $^3$He \cite{Abel1966}. Conventional wisdom holds~\cite{Pethick1988} that for a strong enough repulsive interaction in a given charge or spin channel, ZS excitations are anti-bound states which live outside the particle hole continuum ($v_\zs > v_F$) and appear as sharp peaks in spectroscopic probes, while for attractive interaction they are resonances buried inside the continuum. Possibly the best known example of a resonance is a Landau-overdamped mode near a Pomeranchuk transition ~\cite{Pomeranchuk1959,Abrikosov1975, Landau1980,Baym1991,Nozieres1999,Pethick1988,Oganesyan2001,DellAnna2006,Maslov2010,Watanabe2014,Kiselev2017,
Chubukov2018,Chubukov2019,Sodemann2019}. These qualitative notions are consistent with rigorous results for a 3D FL \cite{Landau1980,Abrikosov1975,Baym1991,Nozieres1999,Pethick1988}.

\begin{figure}
  \centering
  \includegraphics[clip,trim=0 0 0 0,width=0.95\hsize]{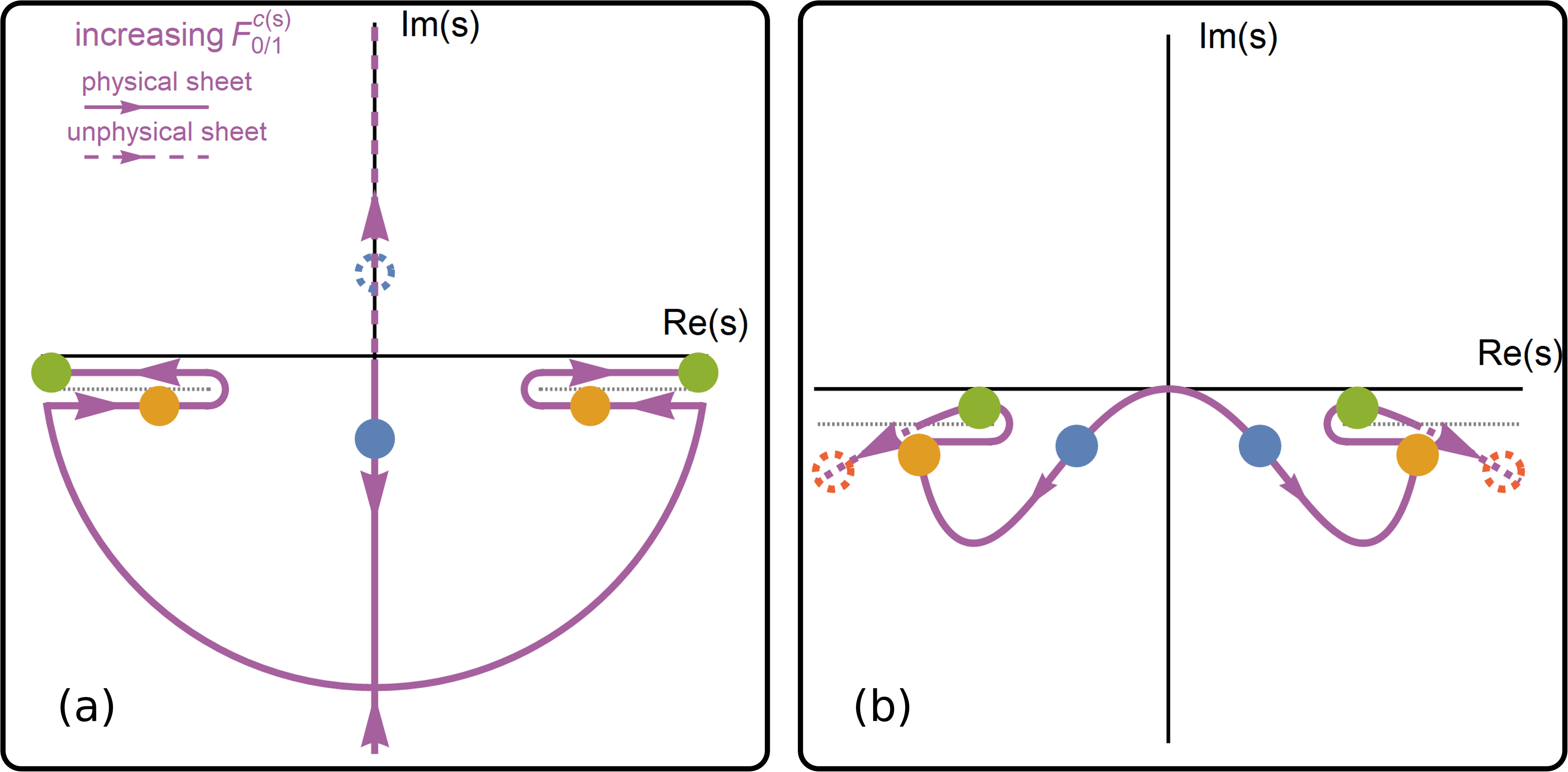}
  \caption{ Trajectories of the poles of $\chi_l^\e(q,\omega)$ on the two-sheeted Riemann surface of complex $s=\omega/v_Fq$.
   Solid
    (dashed) circles denote the poles on the physical (unphysical) Riemann sheet. Solid (dashed) magenta arrows denote the direction of poles' motion on the physical (unphysical) sheet with increasing $F_l^\e$. (a) $l=0$ surface. Blue circle: overdamped ZS mode; yellow circles: hidden mode; green circles: propagating ZS mode.  (b) $l=1$ surface.  Blue circles: damped ZS modes; green: propagating ZS modes; orange circles: mirage modes. For clarity, additional poles on the unphysical sheet are not shown (see  Supplementary Material (SM) \cite{SM}).
    }
  \label{fig:RS-poles}
\end{figure}
In this Letter, we report on two unconventional features of ZS excitations in a clean 2D FL. First,  for  relatively  weak attraction, ZS modes with any angular momentum $l$ are not the expected overdamped resonances but rather sharp propagating modes with $v_\zs > v_F$. However, a spectroscopic probe will not show a peak at $\omega = v_\zs q$. Second, for sufficiently strong repulsion, ZS modes with  $l\geq 1$ appear as sharp peaks in a spectroscopic measurement with $v_\zs > v_F$ but the modes are  not the true poles of the dynamical susceptibility  and as a result  are not the longest lived excitations of the system.
We argue that these two features  come about because  the charge (c)  and spin (s) susceptibilities  $\chi_l^\e(q,\w)$ in  the  angular momentum channel $l$  are   nonanalytic functions of complex $\w$   with   branch points at $\w=\pm v_F q$,   which arise   from  the threshold singularity at the edge of the particle hole continuum. Accordingly, $\chi_l^\e(q,\w)$ is defined  on the complex $\w$ plane  with branch cuts,   located slightly below the real axis in the clean limit (see Fig. \ref{fig:RS-poles}). In 3D, $\chi^\e_l (q, \omega)$ near a branch cut has only a benign logarithmic non-analyticity. In 2D, however, the non-analyticity is algebraic ($\sqrt{x}$). In this situation, the analytic structure of $\chi^\e_l (q,\w)$ is encoded in a two-sheet genus 0 algebraic Riemann surface (a sphere)~\cite{Nehari1952,Farkas1992,Weyl1964}.  It has a physical sheet, on  which $\chi_l^\e(q,\w)$ is analytic in the upper half-plane by causality, and a nonphysical sheet. The ZS modes appear as poles of  $\chi_l^\e(q,\w)$. Both the genus and the number of ZS poles are topological invariants of $\chi_l^\e(q,\w)$,  which remain unchanged as the  poles move on  continuous trajectories over the complex plane. However, to    pass smoothly through a branch cut, a ZS pole must move from the physical to unphysical sheet and vice versa. We show that, for relatively weak attractive interaction,  the propagating pole  is on the physical sheet, but below the branch cut. Consequently, it cannot be analytically extended to the real $\omega$ axis of the physical sheet and does not give rise to a sharp peak in  $\im\chi^\e_l (q, \omega)$  above the continuum.  We label such a mode  as  ``hidden''.  It is  similar to the ``tachyon ghost'' plasmon that   appears     in an ultra-clean  2D electron gas once retardation effects are taken into account \cite{Falko1989,Oriekhov2019}.   For sufficiently weak repulsive interaction in   channels with $l\geq 1$,  the pole is located above the branch cut but, when the interaction exceeds some critical value, the pole moves through the branch cut to the unphysical Riemann sheet.
Although the pole is now below the branch cut, it does gives rise to  a peak in $\chi_l (q, \omega)$ because the pole can be continued back through the branch cut to the physical real axis.
We label such a mode as ``mirage".

\begin{figure}
  \centering
  \raggedright\includegraphics[width=0.9\hsize,trim=0 0 0 0,clip]{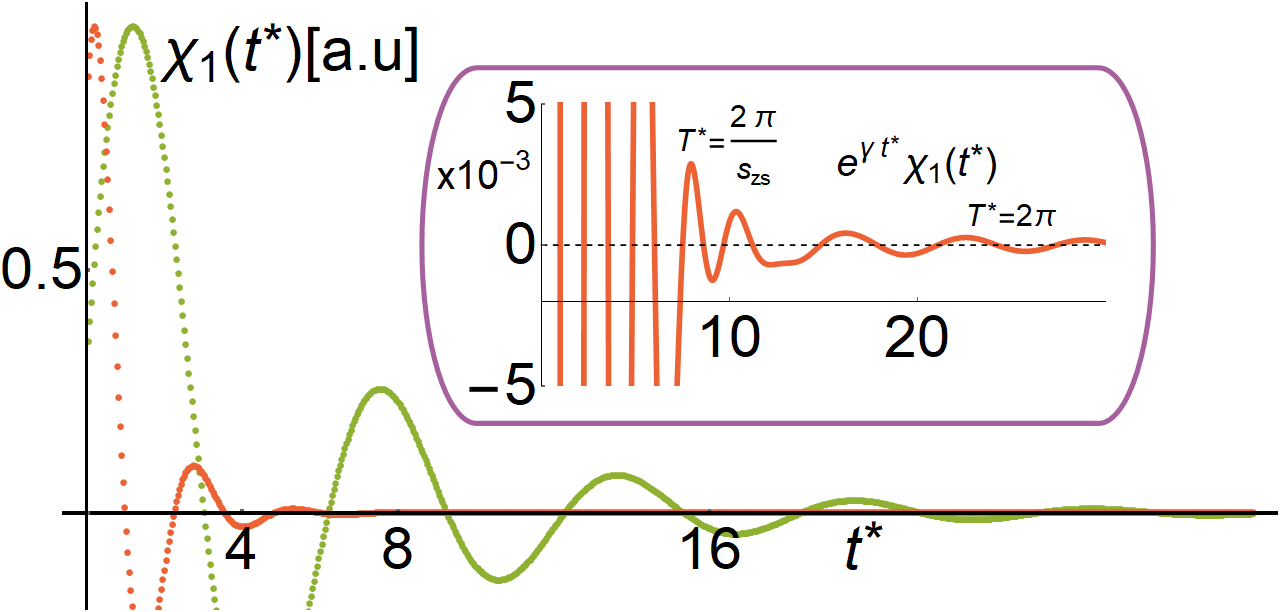}
  \caption{(color online) Time evolution of $\chi_1^\e(t^*)$ for   a conventional ZS mode   at   $F_1^\e = 0.2$   (green) and   a mirage mode at   $F_1^\e=8.0$   (orange). The modes correspond to the green and orange circles in Fig. \ref{fig:RS-poles}b. The conventional mode displays an underdamped behavior with decay constant $\gamma_\zs < \gamma$ and oscillation period $T^* =
    2\pi/s_\zs   < 2\pi$ at all times. The mirage mode decays with $\gamma_\zs > \gamma$ and crosses over to oscillations with period $T^* = 2\pi$ at a crossover time $t_\crs \approx (\gamma_\zs - \gamma)^{-1}$. Inset: a zoomed-in view showing the crossover at $t \sim t_\crs$. $\chi_1^\e(t^*)$ is multiplied by $e^{\gamma t^*}$ to enhance visibility. The disorder strength is $\gamma = 0.2$.}
  \label{fig:l1-time}
\end{figure}

\begin{figure*}
  \centering
  \includegraphics[width=\hsize]{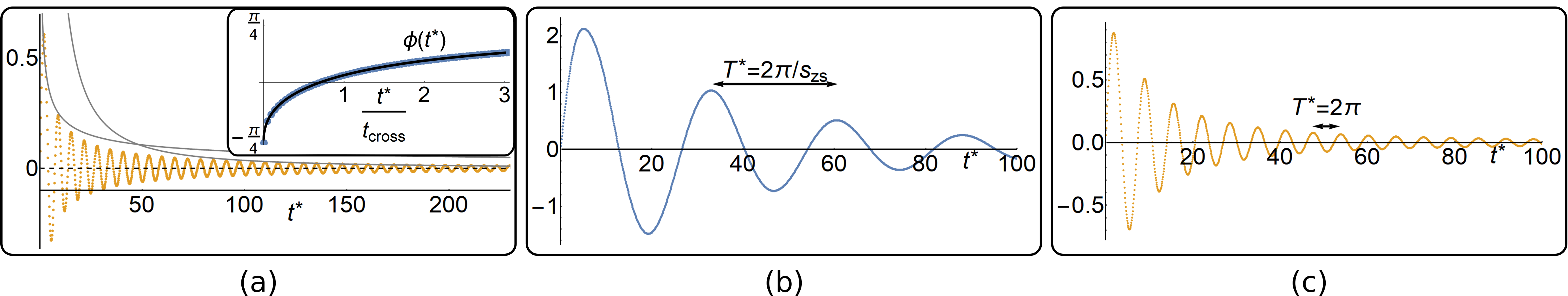}
\caption{(color online) Time dependence $\chi_l^\e(q,t)$ for a system hosting: (a) A hidden mode at $F_0^\e = -0.125$ (yellow circles in Fig. \ref{fig:RS-poles}a). The gray lines show the characteristic power-law decays $\propto t^{-1/2},t^{-3/2}$.     Inset: numerically extracted     variation     of the phase shift     between the two regimes, described in the text (solid), and the analytic prediction (dashed), for $F_0^\e = 0.03$. (b) A damped $l=1$ mode at $F_1^\e = -0.9$ (blue circles in Fig. \ref{fig:RS-poles}b). At      even      longer times (not shown), the period of oscillations approaches $2\pi$. (c) A hidden $l=1$ mode at $F_1^\e = -0.121$ (yellow circles in Fig.
  \ref{fig:RS-poles}b).}
  \label{fig:other-cases}
\end{figure*}

Hidden and mirage
modes cannot be identified spectroscopically by probing $\im\chi^\e_l(q,\omega)$,  as hidden modes do not appear in such a measurement at all, while  mirage modes do appear but cannot be distinguished from conventional modes. We argue, however,   that they  can be  identified by  studying  the transient response of a 2D  FL  in real time,  i.e., by analyzing $\chi^\e_l (q, t)$ extracted from   pump-probe measurements, which have recently emerged as a powerful technique for characterizing and controlling complex materials ~\cite{Krausz2009,Mitrano2016,Giannetti2016,Gandolfi2017,Nosarzewski2017,Nicoletti2016,Hoegen2018,Mitra2019,Gedik2019}. At long times, the response function $\chi^\e_l (q,t)$ is  the sum of contributions from the ZS poles and the branch points. One can readily distinguish a conventional ZS modes from a mirage one via $\chi_l^\e(q,t)$ because a conventional ZS mode is located above the branch cut and decays slower than the branch point contribution, while a mirage mode decays faster.  As a result,  the response of a FL hosting a  mirage  mode undergoes a crossover  from oscillations at the ZS mode frequency to oscillations at the branch point frequency $\omega = v_F q$   at some $t=t_\crs$  (see Fig. \ref{fig:l1-time}). The detection of a hidden mode is a more subtle issue as this mode does not appear on the real frequency axis, and $ \chi^\e_l (q, t)$ at large $t$ always oscillates at $\w = v_Fq$. However, we show that in the presence of the hidden pole the behavior of $\chi^\e_l (q,t)$ changes from $\cos(v_F q t + \pi/4)/t^{1/2}$ at intermediate $t$ to $\cos(v_F q t - \pi/4)/t^{3/2}$ at the longest $t$, and the location of the hidden pole can be extracted from the crossover scale $\bar t_{\crs}$ between the two regimes (see Fig. \ref{fig:other-cases}a).

{\it \bf {Zero-sound modes in 2D.~~~}}
\label{sec:model-results}
A generic bosonic excitation of a FL  with angular momentum $l$ and dispersion $\omega (q)$ is the solution of $\left(\chi_{l}^\e( q,\omega) \right)^{-1} =0$. ZS  excitations are the modes with linear dispersion $\omega = v_{zs}q$ in the limit $q \ll k_F$, where $k_F$ is the Fermi momentum.   The quasiparticle susceptibility at small  $\omega$ and $q$ but fixed $\omega/v_F q =s$ is expressed solely in terms of Landau parameters $F_l^\e$ in the charge or spin sectors ~\cite{Pomeranchuk1959,Landau1980,Abrikosov1975,Baym1991,Pethick1988,Nozieres1999,Chubukov2018,Sodemann2019,Chubukov2019}.
 An explicit form of $\chi_{l}^\e( q,\omega)$ is rather cumbersome but becomes much simpler if one of the Landau parameters, $F^\e_l$, is much larger than the others.  Up to an irrelevant overall factor, for this case we have
\begin{equation}
  \label{eq:chi-qp-def}
  \chi_{l}^\e\left(s\right) \propto   \frac{\chi_{l}(s)}{1+F_l^\e\chi_{l}(s)},
\end{equation}
where $\chi_{l} (s)$ is the quasiparticle contribution from states near the FS, normalized to $\chi_{l}(0)=1$. The general structure of $\chi_{l}(s)$ can be inferred from the particle-hole bubble of free fermions with propagators $G_0(\mathbf{k},\w) =
\left(\w + i\tilde{\gamma}/2 - v_F(|\mathbf{k}|-k_F )\right)^{-1}$ and form-factors $f_l (\theta)$ at the vertices, where $\theta$ is the angle between ${\bf k}$ and ${\bf q}$, $f_0=1$, and $f_l (\theta) = \sqrt{2} \cos l\theta\;
(\sqrt{2}\sin l\theta)$  for the longitudinal (transverse) channels  with $l  \geq 1$.  (The  longitudinal/transverse modes correspond to  oscillations of the FS that conserve/do not conserve its area.) However, to properly specify the position of the pole with respect to the branch cut one must include vertex corrections due to the same scattering processes that give rise to the $i \tilde{\gamma}$ term in $G_0$ (Refs. \cite{Zala2001,Chubukov2019}). This is true even in the clean limit $\tilde{\gamma} \to 0$. To be specific, we assume that extrinsic damping is provided by short-range impurities, and  account for the corresponding vertex corrections in all subsequent calculations.
We study the case
$l=0$ as an example of a hidden mode, and the case
$l = 1$, with $f_l (\theta) = \sqrt{2} \cos{\theta}$, as an example of a mirage mode \cite{footnote1}.
For $l=0$, $\chi_0 (s)$ with vertex corrections due to impurity scattering included is given by~\cite{Zala2001,Chubukov2019}
\begin{equation}
  \label{eq:chi-free-0}
  \chi_{0}(s) = 1 + \frac{i s}{\sqrt{1 - (s + i \gamma)^2} -\gamma},
\end{equation}
where $\gamma = \tilde{\gamma}/v_F q$. Observe that i) $\chi_{0}(s)$ vanishes at $q \to 0$ and finite $\omega$ and $\gamma$, as required by charge/spin conservation, and ii) $\chi_0 (s)$ has branch cuts at $ s = \pm x - i\gamma$, $x>1 $, see Fig. \ref{fig:RS-poles}.
From Eq.~\eqref{eq:chi-qp-def}, the equation  for the pole is $1+F_0^\e\chi_{0}(s)=0$. For $F^\e_0 >0$  and $\gamma\ll 1$,   the two poles are located at $\omega = v_F q \left(\pm s_{zs} -i \gamma_{zs}\right)$, where $s_{zs} = (1+F_0^\e)/\sqrt{1+2F_0^\e} > 1$ and $\gamma_{zs} = \gamma (1+F_0^\e)/(1+2F_0^\e)  < \gamma$. These are conventional ZS poles above the branch cut, which give rise to a peak in $\im\chi_0^\e (q, \omega)$ at $\omega =v_F s_{zs}q$. For  $ -1<F^\e_0 < -1/2$, the two poles are located along the imaginary $s$ axis, one on the physical Riemann sheet, at $s_{zs} = -i (1-|F_0^\e|)/\sqrt{2|F_0^\e|-1}$, and the other on the unphysical Riemann sheet. This is another conventional behavior -- the ZS is Landau overdamped, and at $  F^\e_0 \to -1$ its frequency vanishes, signaling a Pomeranchuk instability~\cite{Pethick1988,Chubukov2019}. The hidden  ZS mode emerges at $-1/2 < F^\e_l <0$. Here the two modes are again located near the real axis, at  $\omega = v_F q \left(\pm s_{ \hzs} -i \gamma_{ \hzs}\right)$, where $s_{ \hzs} = (1-|F_0^\e|)/\sqrt{1-2|F_0^\e|} > 1$ and $\gamma_{ \hzs} = \gamma (1-|F_0^\e|)/(1-2|F_0^\e|)  > \gamma$.
 Since $s_\hzs > 1$, the  ZS  mode is  formally  outside the continuum, i.e.,  it  is an anti-bound state,  even though  the interaction is attractive ($F^\e_0 <0$).   However, because $\gamma_{\hzs} > \gamma$, the pole is located below the branch cut.   Since   a pole cannot pass smoothly through the cut without moving to a different Riemann sheet, a
  hidden pole does not give rise to a peak in $\im\chi^\e (q, \omega)$ at $\omega =  v_F   s_{\hzs} q$. The evolution of the poles with $F_0^\e$ is depicted in Fig. \ref{fig:RS-poles}a.

For $l=1$  one finds:
  \begin{equation}
  \label{eq:chi-free-1}
  \chi_{1}(s) = 1 + 2s^2 \frac{1 + i \frac{s+ i \gamma}{\sqrt{1 - (s + i \gamma)^2}}}{1 -\frac{\gamma}{\sqrt{1 - (s + i \gamma)^2}}}.
\end{equation}
In this case
too,
a hidden pole
exists for attractive interaction,  in the interval $-1/9 < F^\e_1 < 0$. In addition, a new type of behavior emerges for $F^\e_1 >0$.  Namely, $\chi_1^\e$  has  a conventional ZS pole above the branch cut only for
a finite range $0 < F_1^\e < F_1^\mzs$, where $F_1^\mzs = 3/5$ in the clean limit.
At $F_1^\e = F_1^\mzs$
the pole merges with the branch cut and, for larger $F^\e_1$, it moves below the branch cut and, simultaneously, to the unphysical Riemann sheet. We call this pole a ``mirage'' one because although it is located on the unphysical Riemann sheet, it can be connected to the physical real axis through the branch cut. As a result,
 the pole gives rise
 to a sharp
 peak in $\im \chi_1^\e(q,\w)$; however, the width of the mirage mode, $\gamma_\mzs$, is larger than $\gamma$.

{\it \bf {Detection of hidden and mirage modes.~~~}}
We argue that hidden and mirage modes can be observed experimentally by analyzing  the  transient response of a FL which, for an instantaneous  initial perturbation, is described by the susceptibility in the time domain,  $\chi^\e_l (q, t)$. At first glance, it seems redundant to study $\chi_l^\e(q,t)$, which is just a Fourier transform of $\chi^\e_l (q, \omega)$ for real $\omega$, expressed via $\im\chi^\e_l (q,\omega)$ as $\chi^\e_l (q, t>0) = (2/\pi) \int_0^\infty \sin(\omega t)\im \chi^\e_l (q, \omega)$ by causality. A hidden mode does not give rise to a peak in $\im \chi^\e_l (q, \omega)$ for real $\omega$, while the peak due to a  mirage mode is  essentially indistinguishable from that due to a conventional ZS mode. However, we will show below that there are subtle features in $\im \chi^\e_l (q, \omega)$ for  hidden and  mirage modes  that manifest themselves in the time evolution of $\chi^\e_l (q,t)$.

Our reasoning is based on the argument  that $\chi_l^\e(q,t)$ can be obtained by closing the  contour of integration   over $\omega$ on the Riemann surface.
A choice of the  particular contour is a matter of convenience, but a contour can always be decomposed into a part  enclosing the  poles in the lower half-plane (either on the physical or unphysical sheet) and a part connecting the branch points on the Riemann sphere.
For both conventional and mirage  modes the second contribution at long times comes from the vicinity of the branch points and behaves as $\chi^\e_{l}(q, t)\propto \cos(t^*  - \pi/4) e^{-\gamma t^*}t^{-3/2}$, where $t^* = v_F q t$. The pole contribution behaves as $\chi^\e_{l} (q, t) \propto \sin(s_{\text{a}} t^*) e^{-\gamma_{\text{a}} t^*}$, where $\text{a}=\zs,\hzs,\mzs$.  For a conventional ZS mode $\gamma_\zs < \gamma$, and the long-$t$ behavior of $\chi^\e_l (q, t)$ is  dominated by oscillations  at the ZS frequency. For a mirage mode $\gamma < \gamma_{\mzs}$, and the oscillations associated with the mirage mode decay faster than the ones associated with the branch points. We illustrate this behavior in Fig.~\ref{fig:l1-time}, which depicts $\chi^\e_1 (q, t)$ at intermediate and long times for $F_1^\e = 0.2$ and $F_1^\e = 8$, which correspond to the cases of a conventional and  mirage zero-sound  mode, respectively.

For a hidden mode, the situation is more tricky as the pole contribution is cancelled out by a portion of the branch cut contribution and so a hidden pole does not contribute directly to $\chi_0^\e(q,t)$. The only oscillations in $\chi_0^\e(q,t)$ are due to the branch points, with a period $T=2\pi/v_Fq $. However, a more careful study shows~\cite{SM} that in the presence of a hidden pole the branch point contribution undergoes a crossover between two types of oscillations with  the same period:  at intermediate $t$, $\chi^\e_0 (q, t) \propto \cos(t^* + \pi/4)/(t^*)^{1/2}$, while at longer $t$, $\chi^\e_0 (q, t) \propto  \cos(t^* - \pi/4)/ (t^*)^{3/2}$. We illustrate this behavior in Fig.~\ref{fig:other-cases}a.  Note that both the $t$-dependence of the envelope changes and the phase  is shifted  by $\pi/2$. The crossover scale $t^*_{\crs}$ is determined by the position of a hidden pole in relation to the branch point. For small $F_0^\e$ it is just $t^*_\crs  = |s_{\hzs} - (1-i\gamma)|^{-1}$; this relation is verified numerically in the SM~\cite{SM}.  Hence, a hidden pole can be extracted from time-dependent measurements even though it does not show up in  spectroscopic probes.

For completeness, we also briefly discuss the behavior of $\chi^\e_0 (q,t)$ in the range $-1< F^\e_0 <-1/2$, where the pole is Landau   overdamped even in the absence of disorder, i.e., $\omega=-iv_Fq \gamma_{\zs}$ \cite{Chubukov2019}.  In this situation,   dynamics at intermediate $t$ is dominated by a non-oscillatory, exponentially decaying pole contribution, while dynamics at longer $t$ is dominated by algebraically decaying oscillations arising from the branch points, with the  period  $T = 2\pi/(v_F q)$. The crossover time is $t_\crs^{-1} =  (\gamma_\zs-\gamma)^{-1}$ to logarithmic accuracy.  We also  present the  results  for $\chi^\e_1(q,t)$  in two   representative   regimes   of   $F^\e_1 <0$.   As shown in Fig. \ref{fig:RS-poles}b, the $l=1$ poles travel in the complex plane, starting from $\omega = 0$ at the Pomeranchuk instability point $F^\e_1 =-1$ and arriving at the lower edge of the branch cut at $F^\e_1 =-1/9$. Near $F^\e_1 =-1$, the poles  are close to the real axis and, accordingly, $\chi^\e_1 (q,t)$ displays weakly damped oscillations (Fig. \ref{fig:other-cases}b).  When $F^\e_1$ crosses the critical value of  $  -1/9$,     the poles  transform     into     hidden ones, and oscillations     are now     controlled by the     branch points (Fig. \ref{fig:other-cases}c).
As a final remark,  we also verified that  the behavior does not change qualitatively  for a more realistic case when  two Landau parameters, $F_0^\e$ and $F_1^\e$, have comparable magnitudes.

{\it \bf {Summary and discussion.~~~}}
In this Letter, we argued that zero-sound collective excitations in a 2D Fermi liquid have  two unexpected features. First, for any angular momentum $l$ and in some range of a negative Landau parameter $F^\e_l$, a zero-sound mode is not a damped resonance inside a particle-hole continuum, as is the case in 3D, but a propagating mode with velocity larger than $v_F$. In the clean limit, a zero-sound pole of  $\chi^\e_l$ is located arbitrary close to the real axis, but still below the branch cut, which  hides  the pole.  Such a   ``hidden'' mode   does not manifest   itself in spectroscopic probes  but can be identified by transient, pump-probe techniques. Second, for $l \geq 1$ and positive $F^\e_l$ above some critical value, a zero-sound pole moves from the physical Riemann surface to the unphysical one  and becomes a ``mirage'' one.   In this situation, $\im \chi^\e_l (q, \omega)$ still has a peak at the pole frequency in the clean limit.    However, the long-time behavior of $\chi^\e_l (q, t)$ is now determined by the branch points rather than by the pole.

Our work establishes that dynamics of a 2D Fermi liquid, even
of
an
isotropic and Galilean-invariant one, is determined not just by the poles of its response functions, but also by topological properties encoded in the Riemann surfaces defined by those functions. Here we studied the simplest case, where the Riemann surface is a closed sphere. There exist more complex cases, e.g., for two
  bands with different Fermi velocities, $v_{F,1}$ and $v_{F,2}$, there are four branch points in the complex plane, at $\omega =\pm v_{F,1} q, \pm v_{F,2} q$, and the associated Riemann surface is a torus.
 In such cases, one should expect new topological features of zero-sound excitations.

A few  remarks about real systems. First, our results apply to both
neutral and charged FLs, with a caveat that for charged FLs the  $l=0$ charge mode becomes a
plasmon \cite{Lucas2018}.
Second, to observe a zero-sound mode, one either needs to either employ finite-$q$ versions of the pump-probe techniques, e.g. time resolved RIXS \cite{Mitrano2019} and neutron scattering \cite{Granroth2018}, or spatially modulate/laterally confine 2D electrons.
The most readily verifiable prediction is the hidden mode in the spin channel, which
occurs for $0<F_0^a<-1/2$. Previous
measurements on a GaAs/AlGaAs quantum well \cite{Tan2005,Tan2006} indicate
that $F_0^a$ for this system is exactly in the required range.

\begin{acknowledgments}
  We thank M.H. Christensen, A. Kamenev, L. Levitov  and L.P. Pitaevskii for stimulating discussions. This work was supported by the NSF DMR-1834856 (A.K. and A.V.C.), NSF-DMR-1720816 (D.L.M.), and UF DSP Opportunity Fund OR-DRPD-ROF2017 (D.L.M.).
   A.V.C. is thankful to the Aspen Center for Physics (ASP) for hospitality during the completion of this work. ASP is supported by National Science Foundation grant PHY-1607611.
\end{acknowledgments}

\newpage

\section{Supplementary material for ``Hidden and mirage collective modes in two-dimensional Fermi liquids''}

In this Supplementary Material we present the details of our calculations of
the charge/spin susceptibility in the time domain, $\chi_l^\e(q,t)$, and discuss
the analytic structure of the Riemann surface of $\chi_l^\e(q,\omega)$. In Sec. \ref{sec:dynam-susc-chi_l}  we discuss the framework to calculate $\chi_l^\e(q,t)$ for a generic $l$ in the charge or spin channel. In Secs. \ref{sec:l=0-case} and \ref{sec:case-l=1-long} we give detailed derivations of $\chi_l^\e(q,t)$ in the $l=0$ and the $l=1$ longitudinal channels and briefly discuss how these calculations
can be extended to arbitrary $l$.
In Sec. \ref{sec:case-when-f_0e} we show that the results, discussed in the main text, i.e. the existence of conventional, hidden, and mirage poles, also hold when two Landau parameters, $F_0^\e$ and $F_1^\e$, have comparable magnitudes.

Throughout these supplementary notes, we assume an isotropic system, such that at low enough momenta and frequency the fermionic dispersion can be approximated as $\w = \varepsilon_{\mathbf{k}}- \mu \approx v_F (|k|-k_F)$, where $v_F$ is the renormalized Fermi velocity $v_F^{(0)} m /m^*$ and $m^*$ is the Fermi liquid (FL) effective mass. We assume that
single-particle states
are damped by
impurity scattering and that the damping rate,
$\tilde{\gamma}$,
is small compared to Fermi energy.
We also assume that the temperature $T$ is low enough such that the quasiparticle damping rate can be neglected, but
still higher than
the critical temperature of a superconducting (Kohn-Luttinger) instability.

\subsection{
  Dynamical susceptibliities $\chi_l^\e(q,\omega)$ and $\chi_l^\e(q,t)$}
\label{sec:dynam-susc-chi_l}
In this section we provide details of our calculations of the response functions
in the frequency and time domains,
$\chi_l^\e(q,\w)$ and $\chi_l^\e(q,t)$. We assume that typical frequencies and momentum transfers are small,
i.e.,
$q\ll k_F$ and $\w \ll E_F$. In this limit the
response of a FL to a weak external perturbation
comes predominantly
from
quasiparticles near the Fermi surface (FS). The quasiparticle contribution to the dynamical susceptibility was obtained by Leggett back in 1965  (Ref. \onlinecite{Leggett1965}).
To get it diagrammatically, one needs to sum up series of bubble diagrams coupled by quasiparticle interactions. For the case when one Landau parameter dominates, the quasiparticle contribution to $\chi_l^\e(q,\w)$ has the form
\begin{equation}
  \label{eq:chi-qp-def}
  \chi_{\qp,l}^\e\left(q,\w\right)= \nu_F\frac{\chi_l(s)}{1+F_l^\e\chi_l(s)}, s = \frac{\w}{v_F q}
\end{equation}
Here the Landau parameter $F_l$ is the properly normalized  $l$'th moment of the
antisymmetrized four-fermion vertex,  $\nu_F$
is the (renormalized) thermodynamic density of states,
and $\chi_l(s)$ is the retarded free-fermion susceptibility in the $l$'th channel. The subscript $\qp$ makes explicit the fact that this is only the quasiparticle response.
The full $\chi_l^\e(q,\w)$ differs from  (\ref{eq:chi-qp-def}) by an overall factor, which accounts for renormalizations by fermions with higher energies, and also contains (for a
non-conserved
order parameter) an additional term, which comes solely from high-energy fermions \cite{Leggett1965}.
These additional terms are relevant for
the full form of the susceptibility near
Pomeranchuk instabilities towards states
with
special
order
parameter \cite{Kiselev2017,Wu2018,Chubukov2019,Zyuzin2018} but
not for collective modes studied in this paper.
The expression for
the
free-fermion susceptibility $\chi_l(s)$ in the presence of impurity scattering is obtained by (a) evaluating a particle-hole bubble using propagators of free fermions with fermionic frequency $\w$ shifted to  $\w + i
\tilde
\gamma$
and (b) summing up the ladder diagrams for the vertex renormalizations due to impurity scattering. The detailed form of $\chi_l(s)$  depends both on the channel angular momentum $l$ and its polarization (longitudinal/transverse). For a detailed derivation of Eq. \eqref{eq:chi-qp-def} and
explicit
forms of $\chi_l(s)$ we refer the reader to Refs. \onlinecite{Zala2001,Chubukov2018,Chubukov2019}. Here we just state the final results for
$\chi_{\text{qp},l}^\e(s)$
and focus on calculating its time-domain form.
To shorten the notations, henceforth we skip the subindex ``qp" in $\chi_{\qp,l}^\e\left(q,\w\right)$, as we did in the main text.

The retarded time-dependent susceptibility is a Fourier transform of $\chi_{l}^\e(q,\w)$:
\begin{equation}
  \label{eq:chi-qp-t}
  \chi_{l}^\e(q,t) = \int_{-\infty}^{\infty}\frac{d\w}{2\pi} e^{-i \w t}\chi_{l}^\e(q,\w) = v_F q\int_{-\infty}^{\infty}\frac{ds}{2\pi} e^{-i s t^*}\chi_{l}^\e(s),
\end{equation}
where $t^* = v_F q t$.
In physical terms, $\chi^\e_l (q,t)$ describes
a response of the order  parameter in the $l$'th
charge or spin
channel to a  pulse-like excitation of the form $h_le^{-i \q \cdot \mathbf{r}}\delta(t)$.

\begin{figure}
  \centering
  \includegraphics[width=0.7\hsize]{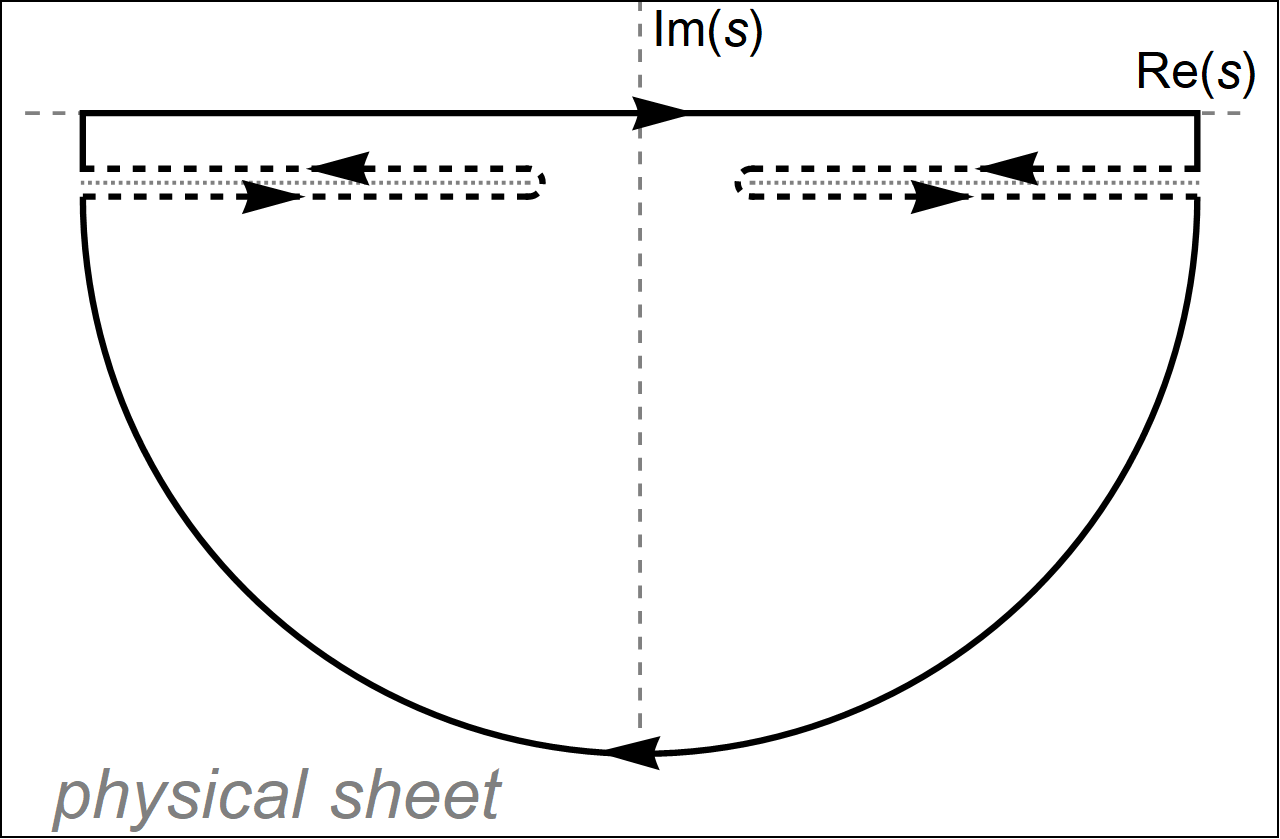}
  \caption{The integration contour over
    (dimensionless)
    complex
    frequency
    $s$ on the physical Riemann sheet.
    \label{fig:cont1}
  }
\end{figure}

To evaluate Eq. \eqref{eq:chi-qp-t}, it is convenient to  close the integration contour in the complex plane. As discussed in the main text, $\chi_l^\e(s)$ has two types of singularities in
complex $s$
plane, both of which contribute to the result of
integration. First, it has a set of poles $s_j$, which can be
either on
the physical or
unphysical sheet. To be concrete, in the subsequent calculations for $l=0,1$ we will label by $s_1$ the pole in the lower-right quadrant of a complex plane of frequency, where $\re s
\geq 0, \im s < 0$. We express the coordinates of the pole $s_1$ as
\begin{equation}
  \label{eq:s1-def}
  s_1 = s_{\text{a}} - i \gamma_{\text{a}},
\end{equation}
where  ${\text{a}} = \zs,\hzs,\mzs$, and the notations are for three different types of the poles corresponding to
a ``conventional'' zero-sound mode (either a propagating one, or a resonance within the particle-hole continuum),
a hidden mode, and a mirage mode, respectively.
These are the same notations that we used in the main text.
To make the text less cumbersome, we will refer to each pole according to the mode it gives rise to, i.e.
we will
call them
a ``conventional pole'',
a``hidden pole'', and
a ``mirage pole".

Second, $\chi_l^\e(s)$ has branch points at $s = \pm 1 - i\gamma$,
where $\gamma=\tilde\gamma/v_Fq$,
and we chose the branch cuts
to run
along the lines  $\pm x - i\gamma, 1< x < \infty$. Because of the sign of the argument of the
exponential function
in Eq. \eqref{eq:chi-qp-t}, the contour must be closed in the lower half-plane
for $t>0$,
so it traces over the branch cuts in the manner shown in Fig. \ref{fig:cont1}.
For $t <0$, the contour must be closed in the upper half-plane,
where $\chi_{l}^\e(s)$ has no singularities and thus $\chi_{l}^\e(q,t<0)=0$
as required by casuality.

The evaluation of the integral over
the
contour in  Fig. \ref{fig:cont1} yields
\begin{equation}
  \label{eq:chi-t-decomp1}
  \chi_l^\e(q,t) = v_F q \chi_l^\e(t^*),\qquad \chi_l^\e(t^*) = \chi^\e_{l,\pole}(t^*) -
  \chi^\e_{l,\bc}(t^*).
\end{equation}
Here $\chi^\e_{l,\pole} (t^*)$ is
a contribution
from the residues of the poles of $\chi_l^\e(s)$ on the physical sheet:
\begin{equation}
  \label{eq:chi-poles-l}
  \chi^\e_{l,\pole}(t^*) = -i\sum_{s_j\in\text{phys.}}e^{-i s_{j} t^*}\text{Res}_{s\to s_{j}}\chi_l(s).
\end{equation}
Since the sum over $s_j$ is restricted to the poles on the physical sheet, it includes  conventional ZS and and hidden poles, but not mirage poles.

The second term in (\ref{eq:chi-t-decomp1}) is the branch-cut contribution
\begin{equation}
  \label{eq:chi-bc-l}
  \chi^\e_{l,\bc}(t^*) = e^{-\gamma t^*}\frac{1}{2\pi}\int_{1}^\infty\left[e^{-i x t^*}\Delta\chi_l^\e(x) + e^{+i x t^*}\Delta\chi_l^\e(-x)\right] dx,
\end{equation}
where $\Delta^\e
\chi_l(x)$ is the
discontinuity of
$\chi^\e_l(s)$ at the branch cut:
\begin{equation}
  \label{eq:delta-chi-l}
  \Delta\chi_l^\e(x) = \lim_{\varepsilon \to 0}\left( \chi_l^\e(x-i\gamma - i\varepsilon) - \chi_l^\e(x-i\gamma + i\varepsilon) \right).
\end{equation}

\begin{figure}
  \centering
  \includegraphics[width=0.48\hsize]{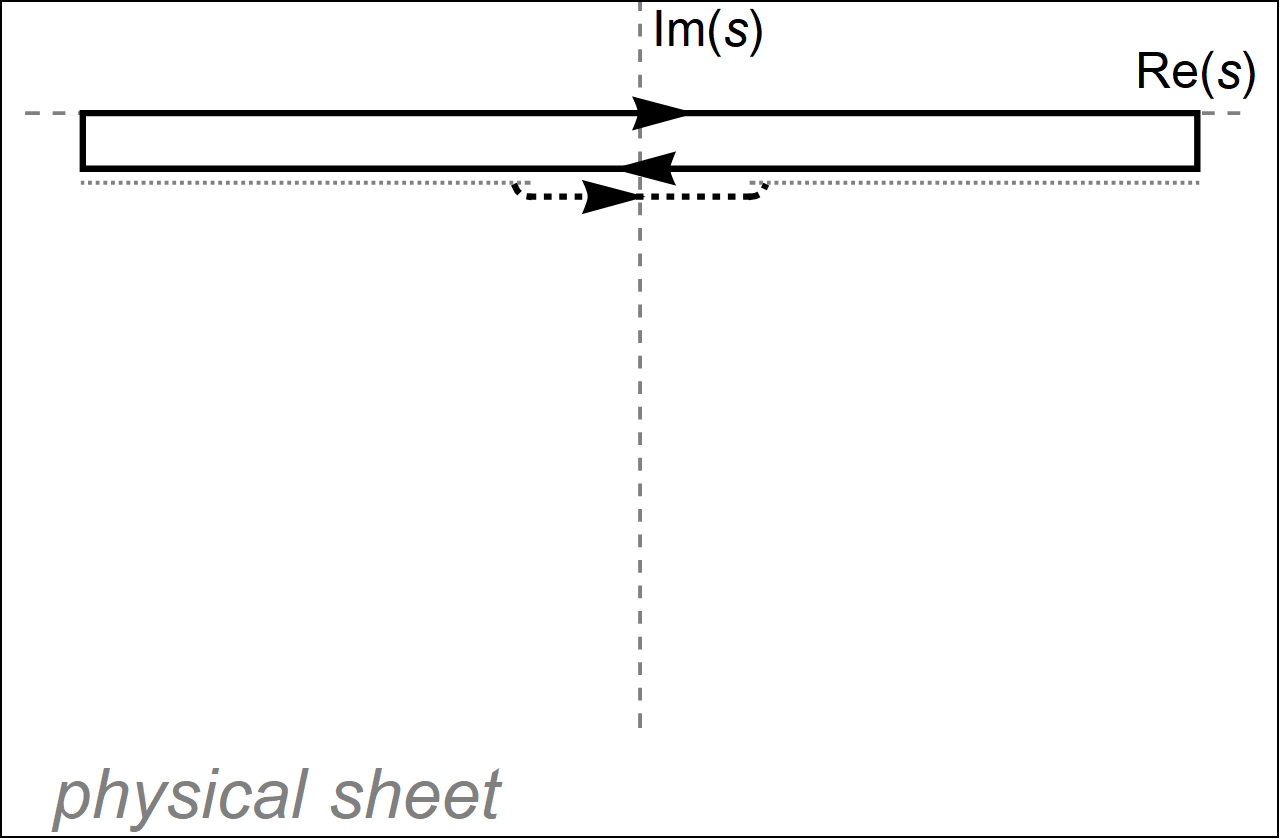}\hfill  \includegraphics[width=0.48\hsize]{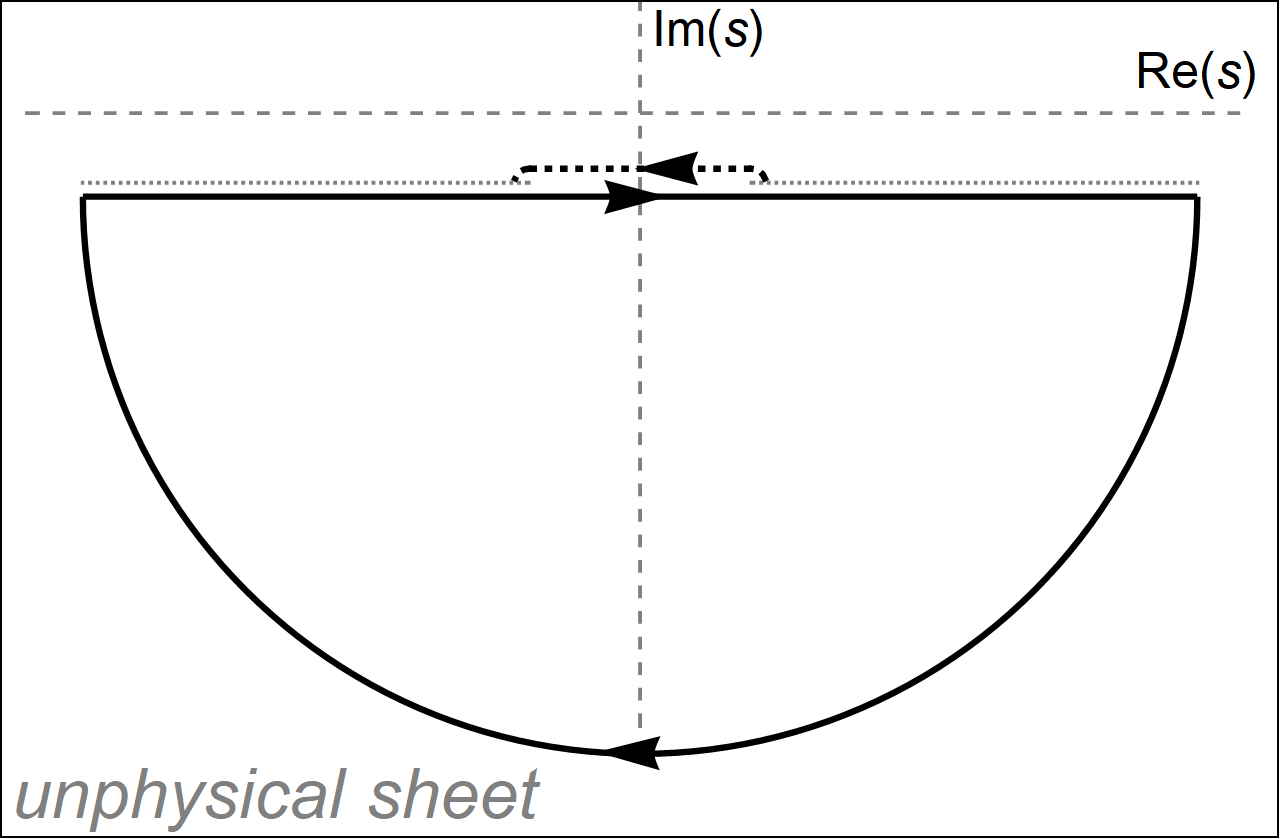}
  \caption{
    Another way to define the integration contour over complex $s$. We added to the integral over real $s$ the integration segments over $s$ immediately above the branch cuts on the physical sheet and immediately below the branch cuts on the unphysical sheet. These  additional integrals then cancel out between the two Riemann sheets. We then added the integral over an infinite semi-circle to the unphysical sheet, and for both sheets added and subtracted the integrals over the range of $s$ between the branch points. The resulting integration contour in each Riemann sheet consists of the closed contour (the solid line) and an additional piece (the dashed line).}
  \label{fig:cont2}
\end{figure}

It is also possible to re-arrange the contour integral into the one depicted in Fig. \ref{fig:cont2}. This is done by (a) closing the integration contour in complex $s$ on the physical sheet along the line $x-i\gamma+i\varepsilon$,  where $\varepsilon$ is infinitesimal and $x=-\infty\ldots\infty$, i.e. along the line which is located right above the branch cuts,  (b) adding an integration contour on the unphysical sheet along the line $x-i\gamma+i\varepsilon, x=-\infty\ldots\infty$, i.e., right below the branch cut, (c) closing this second contour via an infinite half-circle in the unphysical lower half plane, and (d) adding two compensating integration segments along the lines $x-i\gamma -i\varepsilon$, where $-1 \leq x \leq 1$, on the physical sheet, and along $x-i\gamma+i\varepsilon,-1\leq x \leq 1$ on the unphysical sheet
(dashed lines in Fig. \ref{fig:cont2}). Because $\chi_l^\e(s)$ varies smoothly through the branch cuts if one simultaneously move between physical and unphysical Riemann sheets, the integration segments running above and below the branch cuts cancel out.

\begin{figure}
  \centering
  \includegraphics[width=0.5\hsize]{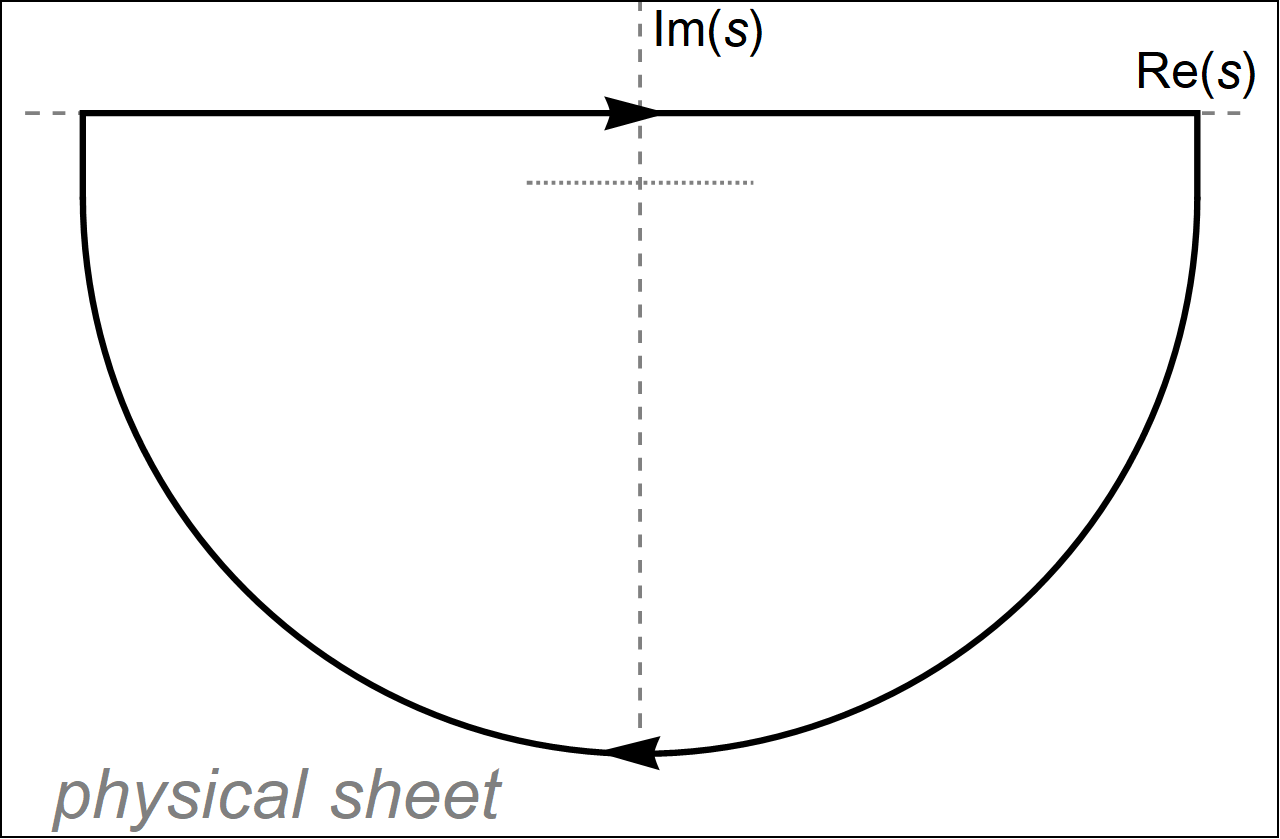}
  \caption{
    Contour of integration over complex $s$ with a branch cut  (dashed line) chosen to run  horizontally  between the branch points at $\mp 1 - i \gamma$.}
  \label{fig:RS2}
\end{figure}

 The evaluation of the integrals again yields an expression of the form of Eq. \eqref{eq:chi-t-decomp1}, but now the sum in  Eq. (\ref{eq:chi-poles-l})
is over the poles on the physical sheet  above the branch cut (i.e., conventional poles with  damping rate $\gamma_\zs < \gamma$), and over mirage poles:
\begin{equation}
  \label{eq:chi-poles-l-2}
  \chi^\e_{l,\pole}(t^*) =
  -i\sum_{s_j\in\text{conv.,mirage}}e^{-i s_{j} t^*}\text{Res}_{s\to s_{j}}\chi_l(s).
\end{equation}
In addition, the second contribution in Eq. (\ref{eq:chi-t-decomp1}) now comes from the difference
between the values of
$\chi_l^\e(s)$ on the two Riemann sheets
rather than from a discontinuity
at the branch cut:
\begin{equation}
  \label{eq:chi-bc-l-2}
  \chi^\e_{l,\bc}(t^*) = e^{-\gamma t^*}\frac{1}{2\pi}\int_{0}^1\left[e^{-i x t^*}\Delta\chi_l^\e(x) + e^{+i x t^*}\Delta\chi_l^\e(-x)\right] dx.
\end{equation}
It can be verified that the integration contour of Fig. \ref{fig:cont2} is equivalent to a contour on the physical sheet, when the branch cut is chosen to run along the line $x - i\gamma, -1 < x < 1$, see Fig. \ref{fig:RS2}.
In this case,
the integral for
$\chi_\bc$
can be understood as running around the circumference of the contour glueing the two Riemann sheets together into a single sphere.

In what follows, we will present calculations using both integration contours, the one in Fig. \ref{fig:cont1} and the one in Fig. \ref{fig:cont2}. Although the result,
of course,
does not depend on the choice of a contour,
some details of the calculation are more transparent when using one contour and some are clearer when using the other.


\subsection{$\chi_l^\e(t^*)$ for $l=0$}
\label{sec:l=0-case}

\begin{figure}
  \includegraphics[width=\hsize]{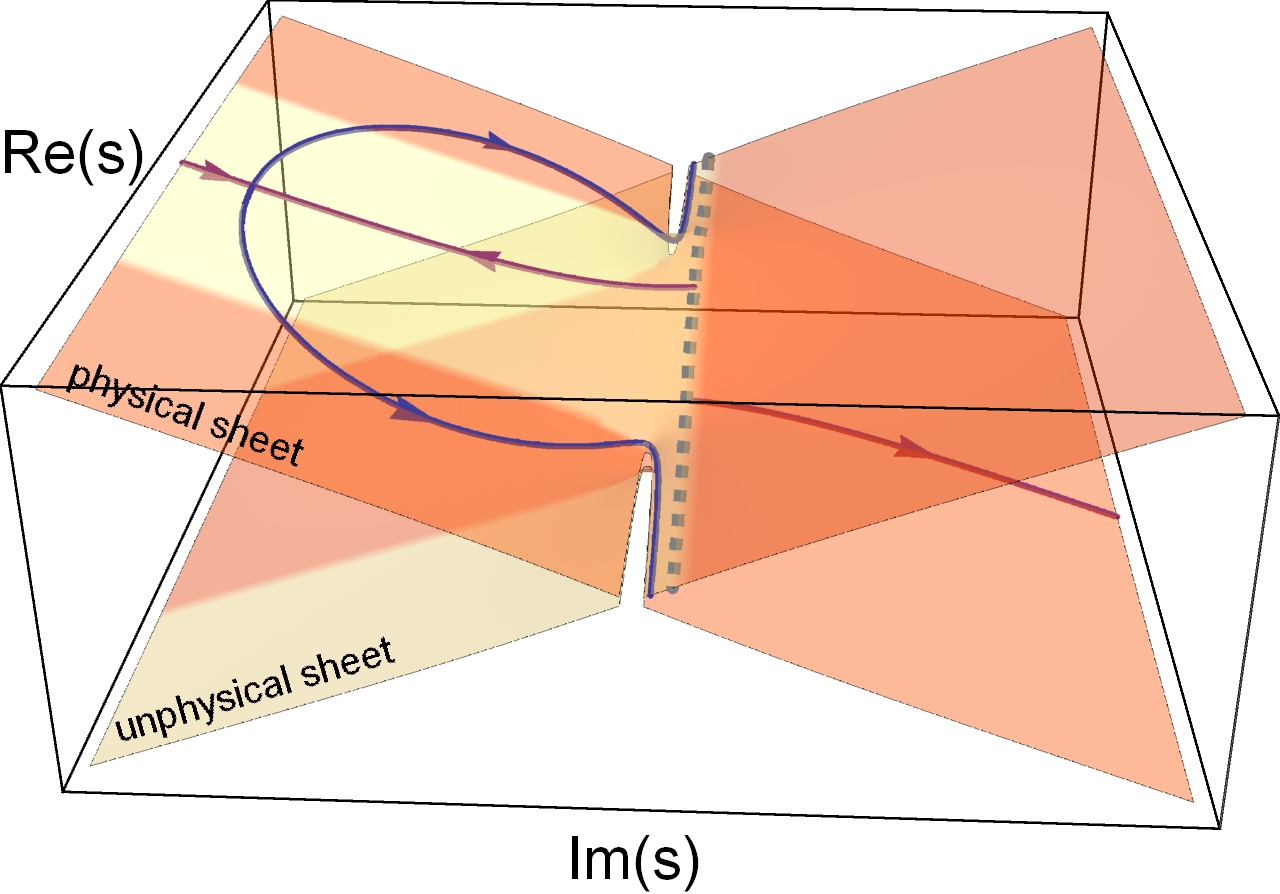}
  \caption{
    (a) A 3D depiction of the pole evolution on the Riemann surface for $l=0$. The figure is obtained by mapping the complex $s$ point of the two Riemann sheets to the 3D set of points $\{\re s, \im s, \pm \re\sqrt{1-s^2}\}$ where $+(-)$ maps the physical (unphysical) sheet to the top (bottom) sheet of the figure. On this representation of the surface, the solid and blue red lines denote the pole evolution with increasing $F_0^\e$. The evolution of the poles begins at the origin of the physical and unphysical sheets at $F_0^\e = -1$. The poles initially move along Im$(s)$  axis down(up) the qphysical(unphysical) sheets. The pole on the unphysical sheet reaches infinity and crosses to the physical sheet at $F_0^\e = -1/2$, and the poles merge and bifurcate at $F_0^\e = -(1-\gamma^2)/2$. The regions with yellow shading denote areas where a pole in $\chi_0^\e(s)$ either on physical, or on unphysical Riemann sheet, gives rise to a peak in $\chi_0^\e(s)$ on the physical real $s$ axis. The areas shaded by peach color are regions where a pole cannot be analytically extended to the physical real axis due to the branch cuts, and $\chi_0^\e(s)$ on the physical real frequency axis has no sharp peaks. We set $\gamma=0.2$ for definiteness.  
    \label{riemannL0}}
\end{figure}

In this section we provide detailed calculations for the case
of $l=0$.
First,
we
use the integration contour in Fig. \ref{fig:cont1} and then the one in Fig. \ref{fig:cont2}.

The free-fermion susceptibility is given by Eq. (2) of the main text
\begin{equation}
  \label{eq:chi-0}
  \chi_{0}(s) = 1 + \frac{i s}{\sqrt{1 - (s + i \gamma)^2} -\gamma}.
\end{equation}
The quasiparticle susceptibility is obtained by plugging $\chi_0$ into Eq. \eqref{eq:chi-qp-def}.
The two  poles of $\chi_0^\e(s)$ are
located
at
\begin{equation}
  \label{eq:s-l-0}
  s_{1,2} = \pm \frac{1+F_0^\e}{1+2F_0^\e} \sqrt{1 + 2 F_0^\e - \gamma^2}  - i\gamma\frac{1+F_0^\e}{1+2F_0^\e}.
\end{equation}
In Fig. \ref{riemannL0} we show a 3D depiction of the poles'
trajectories
on the Riemann surface.
In what follows, we
assume that $\gamma \ll 1$, as we did in the main text.

The discontinuity of $\chi_0(s)$ at the branch cut is
\begin{equation}
  \label{eq:delta-chi-o}
  \Delta\chi_0^\e(x) = \frac{2 \sqrt{x^2-1}(x-i\gamma)}{(1+2F_0^\e)(x - i\gamma - s_1)(x  - i\gamma - s_2)},
\end{equation}
where $s_{1,2}$ are given by (\ref{eq:s-l-0}), see Eq. \eqref{eq:delta-chi-l}.

We obtain $\chi_0(q,t^*)$  for the three  cases shown in Fig. 1a of the main text, i.e.,
for a ZS resonance (an overdamped $l=0$ mode),
hidden mode, and
weakly damped ZS mode.

\subsubsection{ZS resonance, $-1< F^\e_0 <-1/2$}

An overdamped
ZS resonance occurs for $-1 < F_0^\e < -1/2$. The pole contribution can be found directly from Eq.
(\ref{eq:chi-poles-l}).
As follows from Eq. \eqref{eq:s-l-0},
there is
only
one pole in the lower half-plane, at $s_1 = -i \gamma_{zs}$, where
\begin{equation}\label{eq:gamma-zs-conv-0}
  \gamma_{zs} = (1 -|F^\e_0|)/\sqrt{2|F^\e_0|-1}.
\end{equation}
Note $\gamma_{\zs}\gg\gamma$ everywhere but in the narrow vicinity of the Pomeranchuk instability at $F^\e_0=-1$.
Evaluating the residue in Eq. \eqref{eq:chi-poles-l} we obtain
\begin{equation}
  \label{eq:chi-0-t-pole}
  \chi^\e_{0,\pole} (t^*) =  \frac{\sqrt{1+\gamma^2_\zs}}{2|F_0^\e|-1}
  = \frac{|F^\e_0|}{(2|F_0^\e|-1)^{3/2}}e^{-\gamma_\zs t^*}.
\end{equation}
Now we turn to $\chi^\e_{0,\bc} (t^*)$, Eq. (\ref{eq:chi-bc-l}). One can readily verify that at large $t^*$, the leading contribution to the integral in (\ref{eq:chi-bc-l}) comes from the vicinity of the branch point
$s = 1 - i\gamma$. Accordingly, we shift the integration variable in Eq. \eqref{eq:chi-bc-l} to $y=1+x$ and expand the integrand to leading order in $y$. We  obtain
\begin{align}
  \label{eq:chi-0-bcut-conv}
  \chi^\e_{0,\bc} (t^*) &\approx - \frac{2}{ \sqrt{\pi}} e^{-\gamma t^*} \int_0^\infty dy \frac{\sqrt{y}}
                          {(1+2F_0^\e)\sigma_1\sigma_2}
                          e^{-i t^* -i y t^*} + \text{c.c.} \nn\\
                        &= \frac{e^{-\gamma t^*}} {\sqrt{2\pi}(1+2F_0^\e)\sigma_1\sigma_2}
                          e^{-i t^* + i \pi/4}+\cc
\end{align}
where
\begin{equation}
  \label{eq:sigma-def}
  \sigma_{1,2} = s_{1,2} -(1-i\gamma)
\end{equation}
are the pole coordinates  measured from the branch point at $s=1-i\gamma$

Keeping $\gamma$ only in the exponential, we
re-write
Eq. \eqref{eq:chi-0-bcut-conv} as
\begin{equation}
  \label{eq:chi-0-branch-zs-approx}
  \chi^\e_{0,\bc} (t^*) \approx
  e^{- \gamma t^*} \sqrt{\frac{2}{\pi}}\frac{\cos{(t^* - \pi/4)}}{(F_0^\e)^2(t^*)^{3/2}}.
\end{equation}
Comparing $\chi^\e_{0,\pole}$ and $\chi^\e_{0,\bc}$, we see that
at $
F^\e_0 \gtrsim -1$, where $\gamma_{zs} \ll 1$
(but still $\gamma_\zs
>\gamma$),
the pole contribution dominates up to
$t \sim t_\crs$, where
\begin{equation}
  \label{eq:t_crs-conv}
  t_\crs = \frac{3}{2 (\gamma_\zs-\gamma)}\log \frac{(F^\e_0)^2}{(2|F^\e_0|-1)(\gamma_{zs}-\gamma)}
  \gg 1.
\end{equation}
For $t\gg  t_\crs$, the branch-cut contribution becomes the dominant one.
At  $F^\e_0$ not close to $-1$,
$t_\crs\sim 1$.
In this situation,
the branch-cut contribution dominates over the pole one
for all $ t^*\gg 1$.

\subsubsection{Weakly damped ZS mode, $F^\e_0 >0$}

For $F^\e_0 >0$,  ZS  excitations are conventional propagating modes. The time-dependent $\chi^\e_0 (t^*)$ is analyzed along the same lines as for the overdamped
case.
The main difference is that for a propagating mode $\gamma_{zs}  < \gamma$,  and, hence, the pole contribution
remains the dominant one at all times, i.e. there is no crossover to oscillations from the branch point (this incidentally is
indicated by the divergence of $t_\crs$ in Eq. \eqref{eq:t_crs-conv} as $\gamma_\zs$ crosses $\gamma$). The pole contribution is now obtained by summing up the residues of the two poles at $s_{1,2}= \pm s_\zs -i\gamma_\zs$, where  $s_{zs} = (1 + F^\e_0)/(1 + 2 F^\e_0)^{1/2}$ and $\gamma_{zs} = \gamma (1 + F^\e_0)/(1 + 2 F^\e_0) < \gamma$.
Keeping $\gamma$ only in the
exponential,
we find
\begin{equation}
  \label{eq:chi-pole-ZS-prop-exact}
  \chi_{0,\pole(t^*)} = \frac{i \sqrt{s^2_\zs-1}}{(1+2F_0^\e)}e^{ - i s_\zs t^*-\gamma_\zs t^*} + \cc =
  \frac{2 F^\e_0}{(1 + 2F^\e_0)^{3/2}} \sin{s_{zs} t^*} e^{-\gamma_{zs} t^*}
\end{equation}

\subsubsection{Hidden mode, $-1/2 < F^\e_0 <0$}

We next consider the range $-1/2 < F^\e_0 <0$, where the ZS pole is a hidden one: $s_1 = s_\hzs -i \gamma_\hzs$, where $s_{\hzs} = (1- |F^\e_0|)/\sqrt{1-2 |F^\e_0|}$ and $\gamma_{\hzs} = \gamma (1- |F^\e_0|)/(1-2 |F^\e_0|) > \gamma$.
The pole contribution to $\chi^\e_0 (t^*)$ is up to $O(\gamma)$ terms
\begin{equation}
  \label{eq:chi-pole-ZS-prop}
  \chi^\e_{0,\pole} (t^*) =
  -\frac{2 |F^\e_0|}{(1 - 2|F^\e_0|)^{3/2}} \sin{s_{\hzs} t^*} e^{-\gamma_{\hzs} t^*}.
\end{equation}
Note that to get the prefactor right, one has to keep $\gamma$ finite, otherwise the pole and the branch cut would
be at the same depth below the real axis,
and the prefactor in (\ref{eq:chi-pole-ZS-prop}) would be smaller by
a
factor of two because the angle integration around the pole would be only over a half-circle rather than over a full circle.

The branch cut contribution in Eq. \eqref{eq:chi-bc-l} reduces to
\begin{align}
  \chi^\e_{0,\bc}(t^*)  &=
                          \frac{1}{\pi}
                          \frac{e^{-\gamma t^*}}{1 -2 |F^\e_0|}
                          \int_1^\infty dx e^{-i x t^*}
                          \frac{(x-i\gamma)\sqrt{x^2-1}}  {(x - i\gamma - s_1)(x-i\gamma - s_2)} +\cc
                          \label{bb_1}
\end{align}
where now $s_{1,2} = \pm s_\hzs -i\gamma_\hzs$. Evaluating the integral, we find two dominant contributions: one from $x \approx 1$,
i.e., from the vicinity of the branch point,
and
another
one
from $x \approx s_{\hzs}$, i.e., from the vicinity of
the hidden pole (there is only one such term because Re $s_2 <0$).
Accordingly, we write
\begin{equation}
  \chi^\e_{0,\bc}(t^*) = \chi^\e_{0,\bc;a} (t^*) +  \chi^\e_{0,\bc;b} (t^*).
  \label{bb_2}
\end{equation}
To obtain $\chi^\e_{0,\bc;a}$, we expand near $x =s_{\hzs}$ as $x = s_{\hzs} + \epsilon$ and keep the leading terms in $\epsilon$. We obtain
\beq
\chi^\e_{0,\bc;a} (t^*) =  \frac{e^{-\gamma t^*}}{2\pi}
\frac{\sqrt{s^2_{\hzs}-1}}{1 -2 |F^\e_0|}
e^{-i s_{\hzs} t^*} \int_{-\infty}^\infty d \epsilon \frac{e^{-i \epsilon t^*}} {\epsilon + i{\bar\gamma}} +\cc
\label{bb_1_1}
\eeq
where
${\bar \gamma} = \gamma_{\hzs} - \gamma >0$.
The integral in (\ref{bb_1_1}) yields, by Cauchy theorem
\beq
\int_{-\infty}^\infty d \epsilon \frac{e^{-i \epsilon t^*}} {\epsilon + i{\bar\gamma}} =  -2 i \pi e^{- {\bar\gamma}t^*}.
\label{bb_1_2}
\eeq
Substituting into (\ref{bb_1_1}) we obtain
\begin{equation}
  \chi^\e_{0,\bc;a} (t^*)
  =   - 2 \frac{ \sqrt{s^2_{\hzs}-1}}{1 -2 |F^\e_0|} \sin{s_{\hzs} t^*} e^{-\gamma_{\hzs} t^*}
  =  -2 \frac{|F^\e_0|}{(1 -2 |F^\e_0|)^{3/2}} \sin{s_{\hzs} t^*} e^{-\gamma_{\hzs} t^*}.
  \label{bb_1_3}
\end{equation}
Observe that the exponential factor in (\ref{bb_2}) is $e^{- \gamma_
  \hzs
  t^*}$, despite that the overall factor in (\ref{bb_1}) is $e^{-\gamma t^*}$.  The extra factor
$e^{-(\gamma_
  \hzs - \gamma) t^*}$ appears after the integration  in (\ref{bb_1_2}).

Comparing (\ref{eq:chi-pole-ZS-prop}) and (\ref{bb_1_3}), we see that $\chi^\e_{0,\bc;a} (t^*)$ cancels out the pole contribution:
\bea
\chi^\e_{0,\bc;a} (t^*) = \chi_{0,\pole}(t^*).
\label{bb_2_1}
\eea
Because of the cancellation between $ \chi^\e_{0,\bc;a} (t^*)$ and  $\chi_{0,\pole}(t^*)$, there
are
no oscillations in $\chi^\e_0 (t^*)$ with
frequency $s_{\hzs}$, set by the hidden pole.
Note in passing that if we computed $\chi^\e_{0,\bc;a} (t^*)$ strictly at $\gamma =0$, the overall prefactor would be smaller by the factor of two because then $\int_{-\infty}^\infty d \epsilon e^{-i \epsilon t^*}/\epsilon = -i \pi$. The relation $ \chi^\e_{0,\bc;a} (t^*)  = \chi_{0,\pole}(t^*)$ would still hold because the pole contribution  at $\gamma =0$  would also be smaller by a factor of two.

The second term
in Eq.~(\ref{bb_2})
is the contribution from the vicinity of the branch point.  At the largest $t^*$, this contribution has the same form as in Eq. (\ref{eq:chi-0-bcut-conv}):
\beq
\chi^\e_{0,\bc;b} (t^* \to \infty) \approx \sqrt{\frac{2}{\pi}}\frac{\cos(t^*-\pi/4)}{(F_0^\e)^2(t^*)^{3/2}} e^{-\gamma t^*}.
\label{bb_3}
\eeq
However, the full form of $\chi^\e_{0,\bc;b} (t^*)$ is more involved, and
the
$1/(t^*)^{3/2}$ behavior sets in only
after
some characteristic time
$t_{\crs, 1}$, which
becomes
progressively larger as $|F^\e_0|$ decreases and $s_\hzs$ approaches 1. To see this, we expand
the integrand of
(\ref{bb_1}) in $y =x-1$, but do not assume that $y$ is small compared to $\sigma_\hzs = s_\hzs - 1$.
We obtain, at $t^* \gg 1$
\begin{equation}
  \chi^\e_{0,\bc;b} (t^*) \approx -  \sqrt{\frac{2}{\pi t^*}}
  \frac{\sigma_\hzs}{|F_0^\e|^2}e^{-\gamma t^*}
  e^{-i(t^* + \pi/4)} Z(\sigma_\hzs t^*) + \cc,
  \label{bb_4}
\end{equation}
where $z = -i y t^*$ and
\begin{align}
  Z(a) &=  \frac{1}{\sqrt{\pi}} \int_0^\infty dz \frac{\sqrt{z} e^{-z}}
         {z-i a}\nonumber \\
       & = 1 - \sqrt{-i \pi a} e^{-i a}
         \text{erfc}
         \left(\sqrt{-i a}\right),
         \label{bb_5}
\end{align}
where $\sqrt{-i}$ in (\ref{bb_5}) stands for $(1-i)/\sqrt{2}$.
Note that both $\sigma_\hzs$ and $(F_0^\e)^2$
vanish
in the limit $F_0^\e \to 0$,
but
their ratio remains  finite:
$\sigma_\hzs/(F_0^\e)^2 \approx 1/2$.
At small enough $F^\e_0$, $a = \sigma_h t^*$ can  remain small even when $t^*$ is large. Accordingly, we treat $a$ as a variable which can have any value.
In the two limits $a \gg 1$ and $a \ll 1$ we have
\begin{align}
  \label{bb_6}
  Z(a) \approx
  \left\{\begin{array}{ll}
           1, & a \ll 1\\
           \frac{i}{2 a}, & a \gg  1.
         \end{array}\right.
\end{align}
Accordingly, in the two limits $\chi^\e_{0,\bc;b} (t^*)$ behaves as
\begin{align}
  \label{bb_7}
  & \chi^\e_{0,\bc;b} (t^*)  \propto
    \left\{\begin{array}{ll}
             \frac{\cos(t^* + \pi/4)}{(t^*)^{1/2}},  & \sigma_\hzs t^* \ll 1\\
             \frac{\cos(t^*-\pi/4)}{\sigma_\hzs(t^*)^{3/2}}, & \sigma_\hzs t^* \gg  1.
           \end{array}\right.
\end{align}
We see that both
the  exponent of the
power law decay and the phase of oscillations
vary
between the two regimes. In particular,
the phase
changes by $\pi/2$
between the regimes of $\sigma_\hzs t^* \ll 1$ and $\sigma_\hzs t^* \gg 1$
(up to corrections $O(\gamma)$).
The crossover
between the two regimes
occurs  at $t^* \sim t_{\crs,1}$, where
\begin{equation}
  t_{\crs,1} = 1/\sigma_\hzs = 1/(s_\hzs - 1)
  \label{bb_8}
\end{equation}
is related to the coordinate of the hidden pole. This relation provides
a
way to detect the hidden mode
experimentally,
particularly for small $F^\e_0$, where $s_\hzs - 1 \ll 1$ and $t_{\crs,1} \gg 1$, by
either by looking at the crossover in the power-law decay of $\chi^\e_0 (t^*)$ or
by studying
a variation of
the phase shift.

\begin{figure}
  \centering
  \begin{subfigure}{0.48\hsize}
    \includegraphics[width=\hsize]{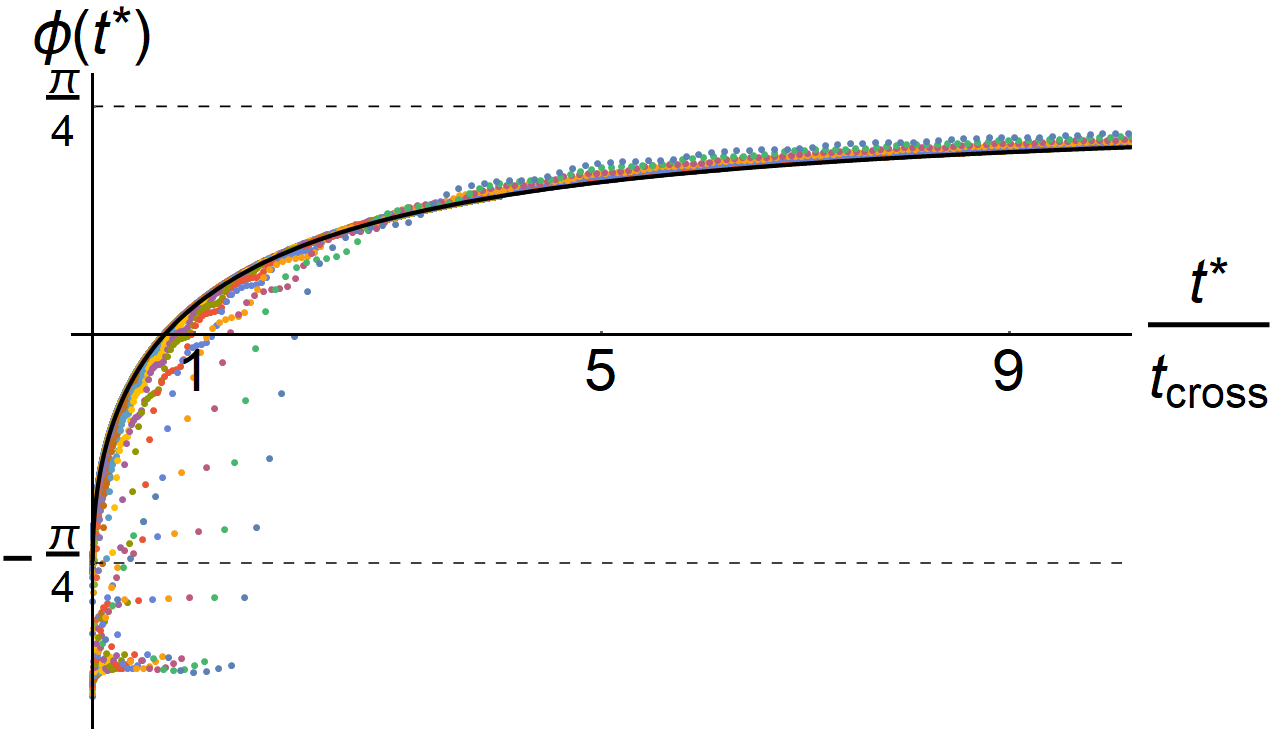}
  \end{subfigure}
  \begin{subfigure}{0.48\hsize}
    \includegraphics[width=\hsize]{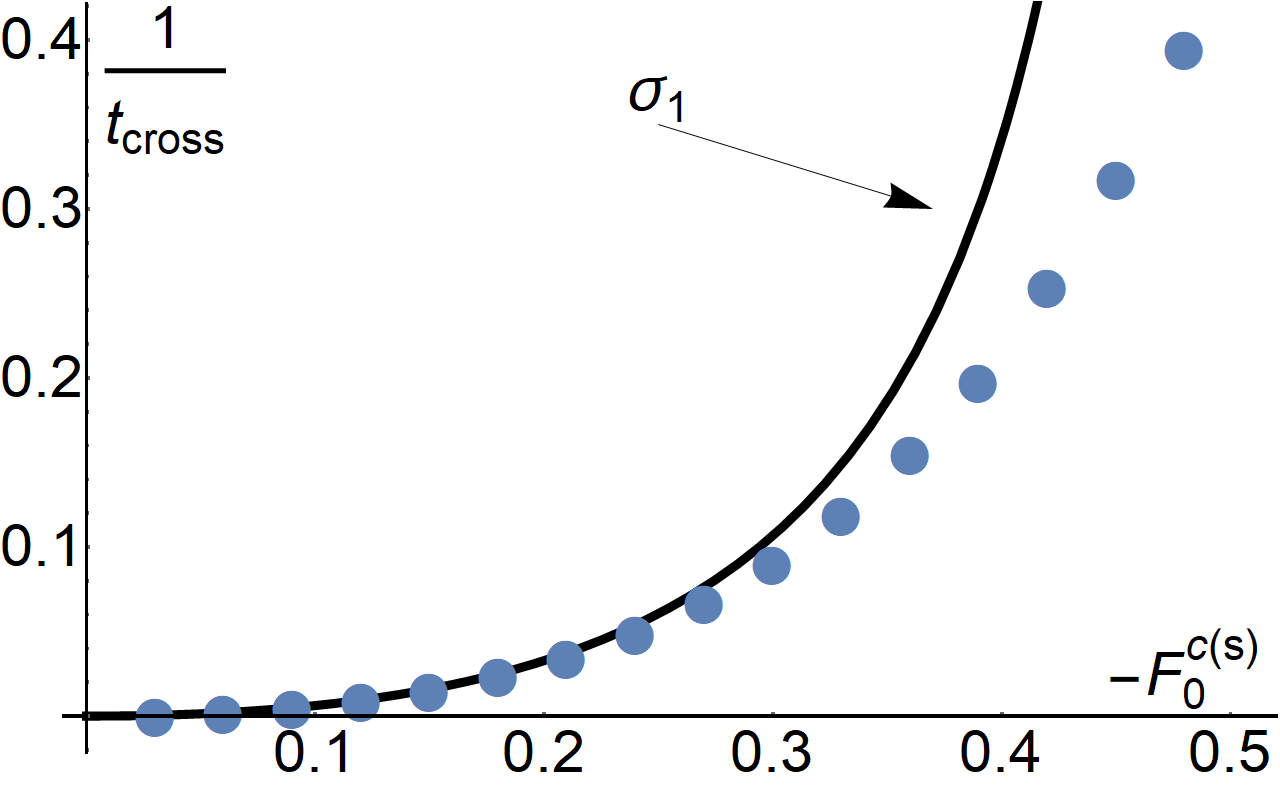}
  \end{subfigure}
  \caption{(a) Evolution of the phase of the oscillations  $\phi(t^*)$ in Eq. (\ref{eq:chi-branch-hidden-0-res})
    with time, for different $F_0^\e = -0.03,-0.06,\ldots -0.48$ (the rightmost blue dots are for $F_0^e =-0.48$).
    Numerical results for $\phi(t^*)$ are plotted as a function of $t^*/t_\crs$. For
    $t > t_{crs}$ the data for different $F_0^\e$ collapse onto a universal curve described by Eq.
    (\ref{bb_4}). (b) Evolution of $t_\crs$ with $F_0^\e$. The black curve is the asymptotic expression in Eq.
    (\ref{bb_8}).}
  \label{fig:phi-evolution-l-0}
\end{figure}

In the intermediate regime
of
$t^* \sim t_{\crs,1}$ (assuming that $t_{\crs,1} \gg 1$)
the susceptibility behaves as
$\chi^\e_0 (t^*) \sim A(\sigma_\hzs t^*) \cos(t^* -\phi (t^*))/(t^*)^{1/2}$. In Fig. \ref{fig:phi-evolution-l-0} we depict $\phi(t^*)$ extracted from numerical evaluation of $\chi_0^\e(t^*)$ for different $F_0^\e$.
\footnote{We fit segments of the data at different $t^*$ onto a trial function $A \cos(t^* - \phi)/(t^*)^\alpha$, where $A,\phi,\alpha$ are fitting parameters. We then fit $\phi(t^*/t_\crs)$ to the prediction of Eq. \eqref{eq:chi-branch-hidden-0-res}.}
The data shows a good collapse of the phase evolution onto a universal
function of
$\sigma_\hzs t^* =
t^*/t_{\crs,1}$,
given by Eqs. (\ref{bb_4}) and (\ref{bb_5}), even for not-too-small $F_0^\e$, and
a
very good agreement
between the numerical value of $t_{
  \crs,1}$
and
the asymptotic expression in Eq. (\ref{bb_8}).

\subsubsection{Calculations using the contour of Fig. \ref{fig:cont2}}
\label{sec:Fig2l0}

We now demonstrate how to evaluate
$\chi_0^\e(t^*)$ in the case of a hidden pole, i.e. at $-1/2< F_0^\e < 0$, using the contour of Fig. \ref{fig:cont2}. The advantage of using this contour is that there is no need to account for
a partial cancellation
between the
pole and brunch-cut
contributions.
 Inspecting
 the integration contours,
 we
 note  that $\chi_{0,\pole}(t^*) = 0$
 because
 there are no poles
 either
  above the branch cuts on the physical sheet
  or below it on the unphysical sheet. We are left only with $\chi_{0,\bc}$, defined in Eq. \eqref{eq:chi-bc-l-2}. We  shift the integration variable in \eqref{eq:chi-bc-l-2} to $y = 1-x$. At $t^* \gg 1$ only small $y$ matter, and one can safely extend the limits of integration to $\pm \infty$.
We then obtain
\begin{equation}
  \label{eq:chi-bc-hidden-0}
  \chi_\bc(t^*) \approx \frac{e^{-it^*}}{2\pi}\int_0^\infty dy \frac{2i\sqrt{2y}(1-i\gamma)}{(1-2|F_0^\e|)(y+\sigma_1)\sigma_2}e^{i y t^*} + \cc
\end{equation}
It is easy to verify that Eq. \eqref{eq:chi-bc-hidden-0} is the analog of Eq. \eqref{bb_1}, up to small corrections due to
$\gamma$.
 The integral in Eq. \eqref{eq:chi-bc-hidden-0} can be
 solved
  exactly
  with the result
\begin{equation}
  \label{eq:chi-branch-hidden-0-res}
  \chi_{0,\bc}(t^*) \approx \frac{e^{-i t^*}i(\sqrt{2i})(1-i\gamma)}{\sqrt{\pi t^*}(1+2F_0^\e)\sigma_2}
  Z(\sigma_\hzs t^*) + \cc
\end{equation}
where $Z(a)$ was defined in Eq. \eqref{bb_5}. This result is the same as in Eq. \eqref{bb_4}, but with corrections due to
finite
  $\gamma$.

We also note in passing that at small $t^* <1$, $ \chi^\e_0 (t^*)$ is linear in $t^*$ for all values of $F^\e_0$. In the limit $\gamma\to 0$ the dependence is given by:
\begin{equation}
  \chi^\e_0 (t^*) = \frac{t^*}{2} \left(1 - \frac{3-2F^\e_0}{24} (t^*)^2 + \cdots \right)
  \label{bb_9}
\end{equation}
At small but finite $\gamma$, the slope at $t^* \to 0$ changes to $\chi^\e_0 (t^*) = (t^*/2) (1 + \gamma \Phi (F^\e_0))$, where $\Phi (-1) =0$ and $\Phi (0-) = 8/(\pi |F^\e_0|)$. For $F^\e_0 =0$, $\chi^\e_0 (t^*) = J_1 (t^*)$, where $J_1$ is a Bessel function.

\subsection{$\chi_l^\e(t^*)$
  in the $l=1$  longitudinal channel}
\label{sec:case-l=1-long}

\begin{figure}[t]
  \centering
  \begin{subfigure}{0.48\hsize}
    \includegraphics[width=\hsize]{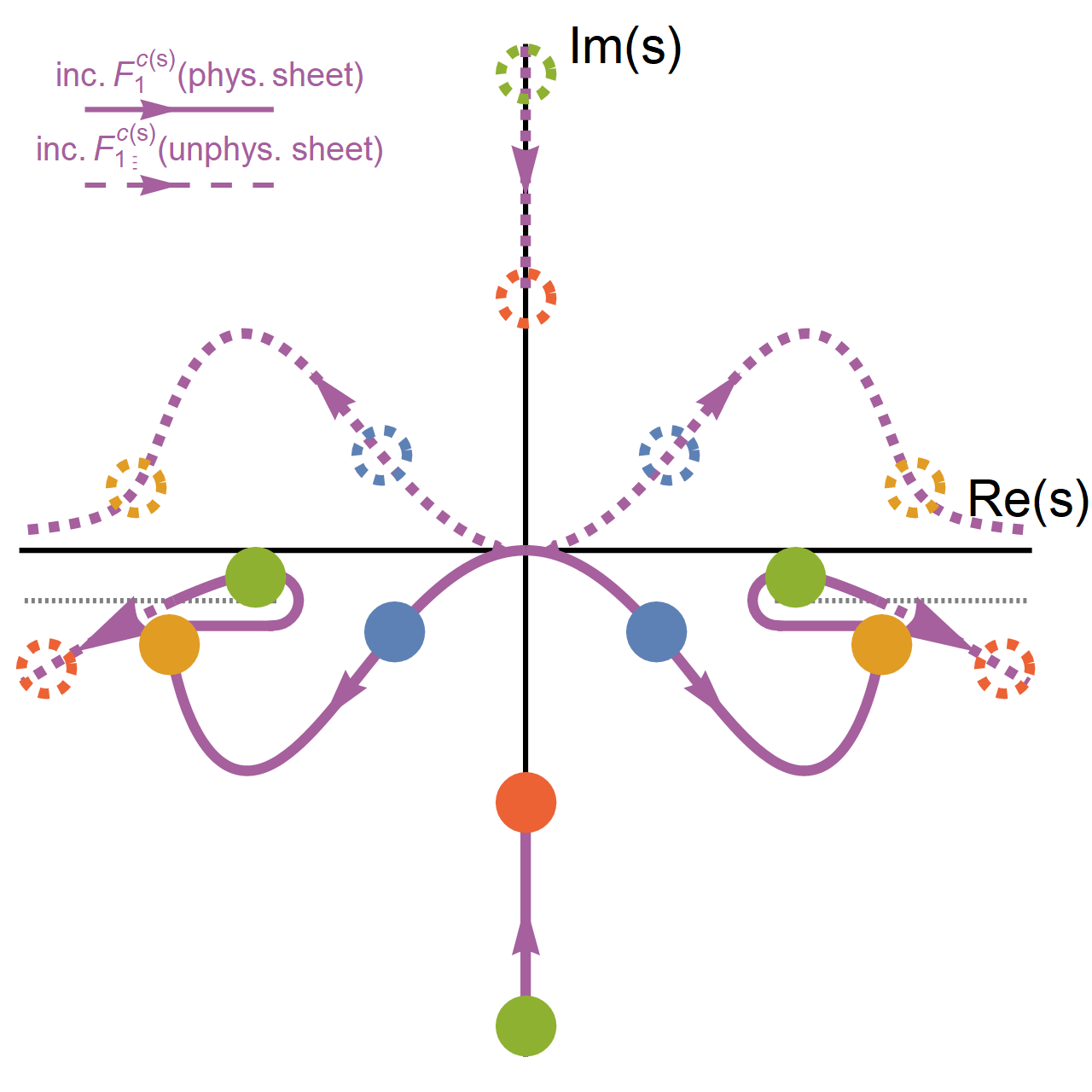}
    \caption{}
  \end{subfigure}
  \begin{subfigure}{0.48\hsize}
    \centering
    \includegraphics[width=\hsize]{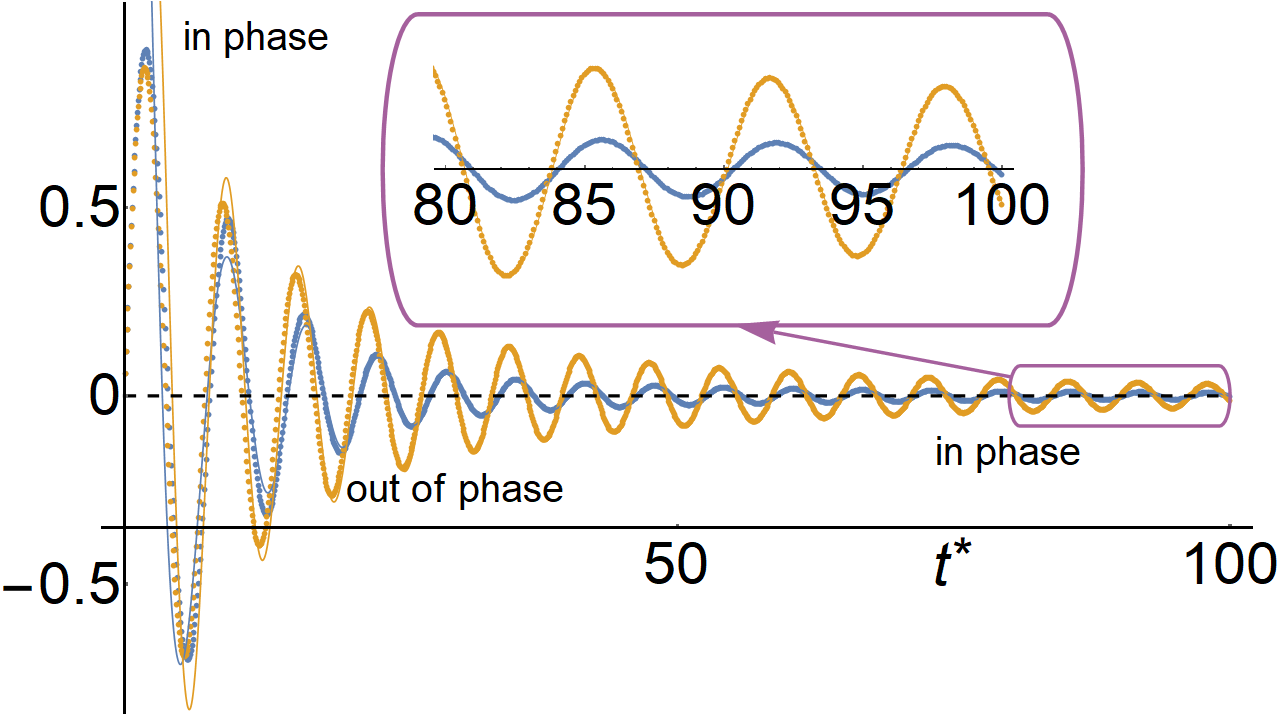}
    \caption{\label{fig:l-1-cross}}
  \end{subfigure}
  \caption{(a)
    A sketch of the trajectories of the poles of $\chi_1^\e(s)$ on the physical and unphysical Riemann surfaces.  Solid
    (dashed) circles denote the poles on the physical (unphysical) Riemann sheet.
    Arrows on
    solid (dashed) magenta
    lines
    denote the direction of poles' motion on the physical (unphysical) sheet with increasing $F_1^\e$. Blue, yellow, green, and orange circles show
     typical positions of
    the poles for the cases of an overdamped ZS mode, a hidden mode, a propagating ZS mode, and a mirage mode, respectively.
    Red circles (solid and dashed) show the positions of additional overdamped ZS modes for $F^\e_1
    > 0$.
    (b)   A crossover in $\chi^\e_1(q,t)$  between the regions dominated by the contributions from the visible  and hidden  poles. The blue (yellow) points denote the numerical result for $F_1^\e = F_1^{\text{vis}} + 0.05$ ($ F_1^{\text{vis}} - 0.05$),
    where  $F_1^{\text{vis}}  = -0.162$, and the solid lines depict the analytical result.
    (The significance of $F_1^{\text{vis}}$ is described in the text around Eq.~\ref{eq:F1-vis-def}.)
    It can be seen that the two traces begin in phase, then move out of phase, and finally become in-phase again. This is an indication that $\chi_1^\e(q,t)$ oscillates at different frequencies that correspond to poles for different $F_1^\e$, until oscillations from the branch points take over  at long times.
  }
  \label{fig:l1-riemann}
\end{figure}

In this section we provide
a detailed derivation of $\chi_1^\e(t^*)$ in the longitudinal channel. The free-fermion susceptibility is
\begin{equation}
  \label{eq:chi-1-free}
  \chi_{1}(s) = 1 + 2s^2 \frac{1 + i \frac{s+ i \gamma}{\sqrt{1 - (s + i \gamma)^2}}}{1 -\frac{\gamma}{\sqrt{1 - (s + i \gamma)^2}}}.
\end{equation}
In the limit $\gamma \to 0$, the pole coordinates are the solutions of
\begin{equation}
  \label{eq:chi-1-poles-clean}
  0 = 4F_1^\e
  s^4
  + \left(1-2F_1^\e - 3(F_1^\e)^2\right)s^2 - (1+F_1^\e)^2.
\end{equation}
This gives 4 poles, which are located  on
both
physical and unphysical sheets. In Fig. \ref{fig:l1-riemann} we present a 2D sketch of the evolution of the four poles on the Riemann surface.
As before,
we  label the pole with Re $s >0$, Im$s >0$
as $s_1$,
We label the pole in the first quadrant of the unphysical sheet
as $s_3$ and define $s_2 = -s_1^*, s_4 = -s_3^*$.
At finite $\gamma$, the expressions for the coordinates of the poles are much more involved, but the number of poles remains unchanged, as does their qualitative behavior.

The discontinuity at the branch cut is
\begin{align}
  \label{eq:delta-chi-1}
  \Delta\chi_1^\e(x)
  = \frac{ \sqrt{x^2-1} (x-i\gamma)^3}{F_1^\e{\displaystyle \prod_{j=1..4}(x-i\gamma-s_j)}}.
\end{align}

Before proceeding to a calculation of $\chi_1^\e(t^*)$ we sketch out the trajectories of $s_{1...4}$ on the physical and unphysical sheets,
see Fig.~\ref{fig:l1-riemann}.
We start with the limit $\gamma \to 0$. The two poles on the physical sheet, $s_{1,2}$, depart from $s=0$ at $F^\e_1 =-1$ and move in the complex frequency plane as $F^\e_1$ increases from $-1$,
until approaching the branch cut
at $F^\e_1 = -1/9$.  For $F^\e_1$ close to $-1$, the poles are almost propagating, and $\gamma_{zs} < \gamma$.
Such poles give rise to oscillations in $\chi^\e_1 (t^*)$ at the pole frequency.  For
$-1/9< F^\e_1 < 0 $, the poles on the physical sheet are hidden. For $0 < F_1^\e < 3/5$,
the poles are conventional ZS poles with $\gamma_\zs < \gamma$. For $ 3/5 < F_1^\e$, the poles move to the unphysical sheet and
become
mirage poles. The two poles on the unphysical sheet, $s_{3,4}$, are
the
mirror images of the
poles on the physical sheet
in the range $-1 < F_1^\e < -1/9$, i.e., $s_3 = s_1^*, s_4 = s_2^*$. In the range $-1/9 < F_1^\e < 0$,
 the two poles move parallel to the real exis, reaching $\pm \infty$ at $F_1^\e = 0$. For positive $F_1^\e$,
the poles $s_3,s_4$ are on the imaginary axis of the lower half plane of the physical sheet, and on the imaginary axis of the upper half-plane of the unphysical sheet. (We recall, that on the Riemann surface the points $\pm \infty, +i\infty$ on the unphysical sheet, and $-i\infty$ on the physical sheet, are identical.) The pole on the physical sheet moves up from $-i\infty$ and the pole on the unphysical sheet moves down from $+i\infty$. At finite $\gamma$, the trajectories are slightly deformed, so that, e.g., $s_{1,2}$ never quite reach the branch cut and $s_{3,4}$ are never true mirror images, but the qualitative behavior remains the same.

We now evaluate $\chi_1^\e(t^*)$. As we did in the $l=0$ case, we first use the contour of Fig. \ref{fig:cont1}. The evaluation proceeds along similar lines as for $l=0$, except for two differences
related, first,
to
the existence of mirage poles, and second,
to
the fact that for some ranges of $F_1^\e$ we need to take into account contributions from all four poles.

\subsubsection{Weakly damped ZS mode, $F^\e_1 \gtrsim -1$}
\label{sec:wdZSl1}

Consider first the limiting case $F^\e_1 \gtrsim -1$. Here $s_1 = s_{\zs} -i \gamma_\zs$, where $s_{\zs} \approx ((1 - |F^\e_1|)/2)^{1/2}$ and $\gamma_{\zs} \approx (1 - |F^\e_1|)/4$. The real part of $s_1$ is much larger than the imaginary
one ($\gamma_\zs\ll s_\zs\ll 1$), i.e., the mode is
underdamped.
The pole and branch contributions
to $\chi^\e(t^*)$
are given by
\begin{align}
  \label{eq:chi-pole-l-1}
  \chi_\pole(t^*) &= \frac{-\sqrt{1-(s_1+i\gamma)^2}s_1^3}{F_1^\e{\displaystyle\prod_{j=2..4}(s_1-s_j)}}e^{-i s_1 t} + \cc, \\
  \chi_\bc(t^*) &\approx \frac{(1-i\gamma)^3}{F_1^\e\sigma_1\sigma_2\sigma_3\sigma_4}\frac{e^{-i t^* + i\pi/4}}{2\sqrt{2\pi}(t^*)^{3/2}} + \cc~,
\end{align}
respectively,
where $\sigma_j = s_j - (1-i\gamma)$,
similar to Eq. \eqref{eq:sigma-def}.
For
$\gamma\to 0$, the pole contribution is
\begin{equation}
  \chi^\e_{1,\pole}(t^*)  \approx \frac{\sin{s_{zs} t^*}}{2 s_{zs}} e^{-\gamma_{zs} t^*}.
  \label{bb_10}
\end{equation}
The branch cut contribution has the same form as in the $l=0$ case, cf. Eq.~(\ref{bb_3}):
\begin{equation}
  \chi^\e_{1,\bc} (t^*) \sim \frac{\cos(t^* - \pi/4)}{(t^*)^{3/2}} e^{-\gamma t^*}.
  \label{bb_11}
\end{equation}
For $F^\e_1 \approx -1$, the
pole contribution
is larger
than the branch-cut one
over a wide range of $t^*$ because the pole contributions
contains a large prefactor $1/s_{zs}$
while the
branch cut contribution is reduced by $1/(t^*)^{3/2}$ at large $t^*$. Still,
at any $|F^\e_1| <1$, intrinsic $\gamma_{zs}$ is finite and by our construction
is
larger than extrinsic
$\gamma$. Then,
at large enough $t^* > t_{\crs,2}$, the branch-cut contribution  becomes larger
than the
contribution from the
pole.
The crossover scale is
\begin{equation}
  t_{\crs,2} \sim \frac{1}{\gamma_{\zs} - \gamma} \log\frac{1}{s_{\zs}(\gamma_{\zs} -\gamma)^{3/2}}.
  \label{bb_12}
\end{equation}
This $t_{\crs,2}$ is
the  $l=1$ analog of $t_\crs$ in the $l=0$ channel, Eq. \eqref{eq:t_crs-conv}.

\subsubsection{
  Hidden pole, $-1/9 < F^\e_1 <0$}

In the hidden pole regime,
which occurs for
$-1/9 < F^\e_1 <0$, the pole contribution is still given by Eq. \eqref{eq:chi-pole-l-1}.  To leading order in $\gamma$, it is
\begin{equation}
  \chi^\e_{1,\pole} (t^*) =- 4 K_\gamma s^2_{\hzs} (s^2_{\hzs} -1)^{1/2}
  \sin{s_{\hzs} t^*},
  \label{bb_14}
\end{equation}
where
\bea K_\gamma =
\frac{1}
{(1-9 |F^\e_1|)^{1/2} (1- |F^\e_1|)^{3/2}}.
\eea
The pole frequency is
\bea s_{\hzs} =
\frac{1-|F^\e_0|}
{8|F^\e_0|}
\left[1+3 |F^\e_0| -\sqrt{(1-|F^\e_0|)(1-9|F^\e_0|)}
\right]^{1/2}.
\eea
In the two limits, $s_{\hzs} =2/\sqrt{3}$ for $F^\e_1 = -1/9$ and $s_\hzs \to 1$
for
$F^\e_1  \to 0$.

To leading order in $\gamma$,
the branch-cut contribution
can be expressed as the sum of the two terms:
\begin{equation}
  \chi^\e_{1,\bc} (t^*) = \chi^\e_{1,\bc;1} (t^*) +\chi^\e_{1,\bc;2} (t^*).
  \label{bb_15_0}
\end{equation}
The first term
contains
the frequency of the pole $s_1$ on the physical Riemann sheet:
\begin{equation}
  \chi^\e_{1,\bc;1} (t^*)  = \frac{2}{\pi} K_\gamma
  \int_1^\infty \frac{dx x^3 \sqrt{x^2-1} e^{-i t^*s}}{(x+i\gamma)^2 -s^2_{1}} + \cc,
  \label{bb_15}
\end{equation}
where we recall that $s_1 = s_{\hzs} - i \gamma_\hzs$ and $\gamma_{\hzs} \geq \gamma$.
The second term
contains the frequency of the pole $s_3$ on the unphysical Riemann sheet:
\begin{equation}
  \chi^\e_{1,\bc;2} (t^*)  = - \frac{2}{\pi} K_\gamma
  \int_1^\infty \frac{dx x^3 \sqrt{x^2-1} e^{-i x t^*}}{(x+i\gamma)^2 -s^2_{3}} + \cc,
  \label{bb_15_1}
\end{equation}
where $s_3 = s'_3  - i \gamma_3$
with $\gamma_3<0$
and
\bea
s'_3 =
\frac{1-|F^\e_0|}
{8|F^\e_0|}
\left[1+3 |F^\e_0| + \sqrt{(1-|F^\e_0|)(1-9|F^\e_0|)}
\right]^{1/2}.
\eea

As for $l=0$, the two largest contributions to  $\chi^\e_{1,\bc;1} (t^*)$ in (\ref{bb_15}) at $t^* \gg 1$ come from $x \approx s_{\hzs}$ and from $x \approx 1$. Accordingly, we further split $\chi^\e_{1,\bc,1} (t^*)$ into
two parts as
$\chi^\e_{1,\bc;1} (t^*) = \chi^\e_{1,\bc;1a}(t^*) + \chi^\e_{1,\bc;1b}(t^*)$.
The first contribution is obtained in the same way as for $l=0$, by expanding in $\epsilon = x - s_{\hzs}$. The result is
\begin{equation}
  \chi^\e_{1,\bc;1a} (t^*)
  =-2 K_\gamma s^2_{\hzs} (s^2_{\hzs} -1)^{1/2}
  \sin (s_{\hzs} t^*)  \left (1 + \frac{\gamma_{\hzs} - \gamma}{|\gamma_{\hzs} - \gamma|} \right).
  \label{bb_15_a}
\end{equation}
Because $\gamma_{\hzs} > \gamma$, the two terms in the last bracket in (\ref{bb_15_a}) are of the same sign
and add up to a factor of 2.
Then
\begin{equation}
  \chi^\e_{1,\bc;1a} (t^*)
  =-4 K_\gamma s^2_{\hzs} (s^2_{\hzs} -1)^{1/2}
  \sin (s_{\hzs} t^*).
  \label{bb_15_1_a}
\end{equation}
This term exactly  cancels out $ \chi^\e_{1,\pole} (t^*)$ from (\ref{bb_14}).
The second contribution,  $\chi^\e_{1,\bc;b}$, yields oscillations with frequency equal to one.
It evinces a crossover from $\chi^\e_{1,\bc;b} \propto \cos(t+ \pi/4)/(t^*)^{1/2}$ behavior at $ t^* < t_{\crs,3}$ to $\chi^\e_{1,\bc;b} \propto \cos(t - \pi/4)/((s_{zs}-1) (t^*)^{3/2})$ behavior at $t^* > t_{\crs,3}$,
where again
\begin{equation}
  \label{eq:t-crs-3}
  t_{\crs,3} = \frac{1}{s_\hzs -1}
\end{equation}
This $t_{\crs,3}$ is the analog of $t_{\crs,1}$ for $l=0$, Eq. \eqref{bb_8}.

The term $\chi^\e_{1,\bc;2} (t^*)$ can  also be split into two contributions, one from $x \approx s'_3$
and another
one
from $x \approx 1$.  Evaluating the first contribution, we find that, up to an overall factor,
\begin{equation}
  \chi^\e_{1,\bc;2a} (t^*) \propto  \sin (s'_{3} t^*)  \left (1 + \frac{\gamma_{3} - \gamma}{|\gamma_{3} - \gamma|} \right).
  \label{bb_15_2}
\end{equation}
Because $\gamma_3 <0$, the second term
in the round brackets
equals $-1$
and cancel the first one.
As a result, there is no $\sin (s'_{3} t^*)$ term in $\chi^\e_1 (t^*)$.  The second contribution,  $\chi^\e_{1,\bc;2b} (t^*)$, has the same structure as
$\chi^\e_{1,\bc;1b} (t^*)$ and just adds up to the prefactor of an oscillation with frequency equal to one.

\subsubsection{Damped ZS mode for $ F^\e_1 \leq -1/9$}
In this section we consider the range of $-1 < F^\e_1 <-1/9$, excluding the immediate vicinity of $-1$, which has been already considered in Sec.~\ref{sec:wdZSl1}.
For $F^\e_1 \lesssim -1/9$ the pole is close to
but somewhat below
the branch cut,
i.e., in our notations this is a weakly damped conventional ZS pole
(by $x\lesssim y$ we mean that $x$ is smaller than $y$ by
an asymptotically small quantity).
Here we have $s_{\zs}  \approx 2/\sqrt{3}, \gamma_{\zs} \approx
\sqrt{3(|F^\e_1|-1/9)/2}$.
Up
to two leading orders in $\gamma_{\zs}$,
the pole contribution is
\begin{equation}
  \chi^\e_{1,\pole} (t^*) =- \frac{3}{2} e^{-\gamma_{\zs} t^*} \left( \frac{\cos{s_{\zs} t^*}}{\gamma_{\zs}} +  3 \sqrt{3} \sin{s_{\zs} t^*} + O(\gamma_\zs)\right).
  \label{bb_16}
\end{equation}
We verified
that both
terms in the pole
contribution
are cancelled out by the corresponding contributions from the branch cut.
The branch cut contribution can again be represented as the sum of two terms, like in (\ref{bb_15_0}), (\ref{bb_15}), (\ref{bb_15_1}), but now $s_3$ is complex conjugate of $s_1$:  $s_3 = s_{\hzs} + i\gamma_{\hzs}$.
The term that cancels (\ref{bb_16}) is obtained by expanding in $\epsilon = x - s_{\hzs}$ and evaluating integrals
up to two leading orders in $\gamma_{\hzs}$.
The cancellation implies that there are no oscillations in $\chi^\e_1 (t^*)$ with frequency  $s_{\zs}$, even when the system is slightly outside the range where the ZS pole is a hidden one.
The remaining contribution from the branch cut has the same form as in other regimes: at largest $t^*$,
\begin{equation}
  \chi^\e_{1,\bc} (t^*)  \propto \frac{\cos(t^* - \pi/4)}{(t^*)^{3/2}}.
  \label{bb_17_1}
\end{equation}

We now study the crossover from the behavior at $F_1^\e \lesssim -1/9$, where we just found that the pole contribution is cancelled by the contribution from the branch cut,
to the  behavior at $F_1^\e \gtrsim -1$, where we
found
earlier
that there is no such cancellation. As $F_1^\e$ decreases, the trajectory of $s_1$ evolves in the complex plane, mirrored by the other $s_{2..4}$. During this evolution, $\gamma_\zs$ is finite but numerically small.  For this reason, below we restrict ourselves to the leading contribution in $\gamma_\zs$.

Within this approximation, the pole contribution is the
first
term in (\ref{bb_16}).
For the branch cut contribution we find, not requiring $s_\zs$ to be close to $2/\sqrt{3}$,
\begin{equation}
  \chi^\e_{1,\bc} (t^*)  =  -  \frac{3e^{i s_\zs t^*}}{4\pi}
  \int_{1-s_\zs}^{\infty} dx \frac{e^{-i x t^*}}{x^2+\gamma_\zs^2} \sqrt{\frac{s_\zs - 1 + x}{s_\zs-1}} + \cc
  \label{bb_17}
\end{equation}
For $s_{zs} <1$, the lower limit of the integral is
positive.
This happens when
\begin{equation}
  \label{eq:F1-vis-def}
  F_1^\e \leq F_1^{\text{vis}},
\end{equation}
where $F_1^{\text{vis}}=-0.162$.
In this range of
$F_1^\e$, one can safely set $\gamma_{zs}$ to zero -- the integral does not diverge. As
a
consequence, $\chi^\e_{1,\bc;1} (t^*)$ does not contain the factor $\propto \gamma_\zs^{-1}$ and cannot cancel  $\chi^\e_{1,\pole} (t^*) \propto \cos(s_{\zs} t^*)/\gamma_{\zs}$ in (\ref{bb_16}). The leading contribution to the integral in (\ref{bb_17}) comes from $x \approx 1-s_{zs}$, and the integration yields
\begin{equation}
  \chi^\e_{1,\bc} (t^*)  \propto \frac{\cos(t^* - \pi/4)}{(t^*)^{3/2}},
\end{equation}
as in (\ref{bb_17_1}).
We see that the behavior  of $\chi^\e_1(t^*)$
is qualitatively the same as for $F \geq -1$: the pole contribution yields oscillations with frequency $s_{zs}$ and remains the largest contribution to $\chi^\e_1 (t^*)$  up to $t^* \sim t_{\crs,2}$. At  $t^* >t_{\crs,2}$, the branch cut contribution becomes the largest one and $\chi^\e_1 (t^*)$
oscillates at
the (dimensionless) frequency equal to one.

However, when $s_{zs} >1$, which happens
for $F_1^{\text{vis}}<F^\e_1<-1/9$,
the  lower limit of integration in Eq. \eqref{bb_17} is
negative,
and
the integral
contains a singular contribution from $x
\to
0$. Using
\begin{equation}
  \int_{-\infty}^\infty
  \frac{e^{-i x t^*}}{x^2 + \gamma^2_{zs}} = \frac{\pi}{\gamma_{zs}} e^{-\gamma_{zs} t^*},
  \label{bb_19}
\end{equation}
we find that this singular piece
cancels out the contribution from the pole. Evaluating the other relevant contribution from $x \approx 1-s_{zs}$, we find
\begin{align}
  \chi^\e_1 (t^*)
  &= -\frac{3}{2 \sqrt{\pi} (s_{zs}-1)^{5/2}}  \frac{\cos((t^* - \pi/4))}{(t^*)^{3/2}}.
    \label{bb_20}
\end{align}
This result is valid for $t^* |s_\zs-1| \gg 1$.  The $\cos(t^* - \pi/4)/(t^*)^{3/2}$ is precisely the expected time dependence
for the case
when the contribution to $\chi^\e_1 (t^*)$ comes solely from the end points of the branch cut.

We see therefore that oscillations with frequency $s_{zs}$ exist  as long as
$F^\e_1 <F_1^{\text{vis}}$. For
$ F_1^{\text{vis}}<F^\e_1<-1/9$
only oscillations, coming from the branch points, with frequency equal to one are present.

In the analysis above we expanded in $\gamma_{\zs}$, i.e.,
we
assumed that the damping remains small in the crossover regime around $
F_1^{\text{vis}}
$.  The approximation of small $\gamma_{\zs}$ would be rigorously valid if
the pole trajectory in the complex plane would remain close to the real axis for all $-1 < F^\e_1 < -1/9$.  In that case we would expect oscillations to persist for a long time, both at $F^\e_1 < F_1^{\text{vis}}$ and at $-1/9 < F^\e_1 < F_1^{\text{vis}}$.
For $F_1^\e < F_1^{\text{vis}}$ oscillations would occur with frequency
$s_{\zs}$ at intermediate $t^*$ (but still $t^* \gg 1$) and with frequency equal to one at  even larger $t^*$. For $F_1^{\text{vis}} <F_1^\e$ oscillations would occur with frequency equal to one
at all $t^* \gg 1$. We see therefore that the branch contribution ``eats up'' the pole contribution once the coordinate of the pole in the complex plane moves to below the branch cut.
In reality, $\gamma_{\zs}$ is small (or order $\gamma$) near $F^\e_1  =-1$ and $F^\e_1 = -1/9$, but is of order one at $F^\e_1 \sim F_1^{\text{vis}}$.
In this situation, the crossover between the behaviors at $F^\e_1
\gtrsim
-1$ and $F^\e_1
\lesssim
-1/9$ is expected to be obscured by damping.  Nevertheless, in numerical calculations, we do see indications of the
crossover
in the behavior of $\chi^\e_1 (t^*)$, when
$F^\e_1$ is varied around $F_1^{\text{vis}}$, see Fig. \ref{fig:l1-riemann} b
and its caption.

\subsubsection{Calculations using the contour of Fig. \ref{fig:cont2}}

We now obtain the same results by using the integration contour of Fig. \ref{fig:cont2}. Again, the use of this contour will allow us to avoid canceling out pole and branch contributions.
It also allows one
to
see more transparently how the poles on the unphysical sheet contribute to the dynamics.
We study both the regime
of hidden poles
and the crossover regime
between
$F^\e_1=-1$ and
$F_1^\e
-1/9$.
For consistency we define $s_1 = s_\zs -i\gamma_\zs$
and $\sigma_\zs = s_1 - (1-i\gamma)$.
With the contour of Fig. \ref{fig:cont2}, the pole contribution is zero
for the same reason as for the $l=0$ case (cf. Sec.~\ref{sec:Fig2l0}),
and the dynamics
is determined entirely
by the branch-cut contribution, which is
given by
\begin{equation}
  \label{eq:chi-bcut-l-1-hidden}
  \chi_{\bc}(t^*) = \frac{e^{-i t^*}}{2\pi}\int_{0}^1e^{i y t^*}\Delta\chi_1^\e(1-y) dy + \cc,
\end{equation}
where we used Eq. \eqref{eq:chi-bc-l-2} and shifted the integration
variable
via $y = 1-x$.
To proceed further, we infer from Eq. \eqref{eq:delta-chi-1} that the $y$ integral is dominated by the region $y \ll |\sigma_i|$, i.e., by whichever pole is nearest to the
branch point, see Eq. \eqref{eq:sigma-def}. In our notations, it is $\sigma_1\equiv\sigma_\zs$. For $|\sigma_\zs| \ll 1$ we may expand the integral in small $y$ and extend the integration limits to infinity. This  yields
\begin{equation}
  \label{eq:chi-bcut-l-1-hidden-approx}
  \chi_{1,\bc}(t^*) \approx \frac{\sqrt{2}i(1-i\gamma)^3}{2\pi F_1^\e}e^{-i t^*}\int_{0}^\infty\frac{\sqrt{y}  e^{i y t^*}}{{\displaystyle \prod_{j=1..4}(y+\sigma_j)}}~dy + \cc
\end{equation}
First,
we consider
the situation when $F_1^\e < 0$ and $|F_1^\e| \ll 1/9$, i.e., when $s_{1,2}$ reside below the branch cut (see Fig. \ref{fig:l-1-cross}) and are close to the branch point. In this situation
$|s_{3,4}|\gg 1$ and the $y$ dependence in the $(y+\sigma_3)(y+\sigma_4)$
factor
in Eq. \eqref{eq:chi-bcut-l-1-hidden-approx}
can be neglected. Then Eq. \eqref{eq:chi-bcut-l-1-hidden-approx} is identical to Eq. \eqref{eq:chi-bc-hidden-0}, up to unimportant constant factors, i.e., the hidden pole behavior for $l=1$ is the same as for $l=0$. Next, we consider the situation when $F_1^\e$ decreases
 and becomes smaller than $-1/9$.
We evaluate the integral in Eq. \eqref{eq:chi-bcut-l-1-hidden-approx} exactly by
contour integration in the first quadrant of complex $y$ and obtain
\begin{equation}
  \label{eq:chi-bcut-l-1-2}
  \chi_{1,\bc}(t^*) \approx \frac{\sqrt{2}i(1-i\gamma)^3}{2\pi F_1^\e}e^{-i t^*}\sum_{j=1..4}A_j\mathcal{Z}(\sigma_j,t^*) + \cc,
\end{equation}
where $A_j=\sum_{i\neq j}(\sigma_i-\sigma_j)^{-1}$ are the partial fraction decompositions of $\prod_j(x+\sigma_j)$, and
\begin{equation}
  \label{eq:Phi-t-def}
  \mathcal{Z}(\sigma,t) = \int_0^\infty dx e^{i x t}\frac{\sqrt{x}}{x+\sigma} dx  = \Theta(-\re\sigma)\Theta(-\im\sigma)2\pi i\sqrt{-\sigma}e^{-i\sigma t} + e^{i\pi/4}Z(\sigma t),
\end{equation}
where $Z(a)$ was defined in Eq. \eqref{bb_5} and $\Theta(a)$ is the Heaviside function.
\footnote{Note that since $s_{2,3}$ are not near the branch point at $1-i\gamma$, they have $\sigma_j \approx -2$ while the integral is dominated by the region $y \sim |\sigma_1|,|\sigma_4|$. However, their contribution is included in the complex conjugate term in $\chi_{1,\bc}$}

Equations \eqref{eq:chi-bcut-l-1-2} and \eqref{eq:Phi-t-def} are applicable in both the hidden pole regime and the crossover regime, as long as $|\sigma_1| \ll 1$.
Let us examine them in the crossover regime.
Although the sum in Eq. \eqref{eq:chi-bcut-l-1-2} is over all four poles, the Heaviside functions in Eq. \eqref{eq:Phi-t-def} are nonzero only for $s_1$.
It can be verified that the sudden appearance of the pole contribution
for $s_1$
is mirrored by a jump in $\sum_jA_jZ(\sigma_jt)$, so that the crossover is actually smooth - the pole progressively ``emerges'' from behind the branch cut. This behavior is the analog of the progressive ``eating up'' of the poles that we obtained via integration over the contour of Fig. \ref{fig:cont1}, see Eq. \eqref{bb_17}.

To obtain a qualitative understanding of how the poles emerge, we expand Eqs. \eqref{eq:chi-bcut-l-1-2} and \eqref{eq:Phi-t-def} in small $\gamma_\zs - \gamma$. This approximation is analogous to the one we made above when studying the crossover using the contour of Fig. \ref{fig:cont1}, i.e. of keeping only the leading contribution in $\gamma_\zs$. Using our results for the contour of Fig. \ref{fig:cont2}, the only necessary step is to take the limit $\im \sigma_j \to 0$ in Eqs. \eqref{eq:chi-bcut-l-1-2} and \eqref{eq:Phi-t-def}, which yields,
\begin{equation}
  \label{eq:chi-branch-artificial}
  \chi_{1,\bc}(t^*) \propto -\Theta(1-s_\zs)2\pi\sqrt{1-s_\zs}e^{-i s_\zs t^*} - e^{-i t^*+i\pi/4}\frac{\sqrt{-i\pi^2(s_\zs-1)}}{4(s_\zs-1)} + \cc,
\end{equation}
i.e. oscillations at a frequency $s_\zs\neq 1$ begin to emerge precisely when $s_\zs < 1$. Eq. \eqref{eq:chi-branch-artificial} is valid when $|(1-s_\zs)t^*| \ll 1$.

\subsubsection{Mirage poles}

Finally, we discuss the mirage poles.
For $0<F_1^\e<3/5$, the conventional ZS pole $s_1$
is located outside particle-hole continuum, and its position in the lower half-plane of frequency is between the real frequency axis and the branch cut, i.e., $\re s_1 >1$
and $-\gamma <$ \im $s_1 <0$.
At $F^\e_1 =3/5$, $\im s_1$ becomes equal to $\gamma$, and for larger $F^\e_1$, the pole moves to
the unphysical Riemann sheet,
i.e.  in our notations it becomes a mirage pole (see Ref. \cite{Chubukov2019}).

As before, we first compute $\chi^\e_1 (t^*)$ using the integration contour in Fig. \ref{fig:cont1}.
Because there are no poles on the
physical
Riemann sheet
for $F_1^\e>3/5$,
the whole contribution comes from the branch cut: $\chi_1^\e(t^*) = - \chi^\e_{1,\bc}(t^*)$.  The integral over the branch cut has two relevant contributions.
The first one,
$\chi^\e_{1,\bc;a_m}(t^*)$, comes from the vicinity of branch points.  This contribution is computed in the same
way
as
the
analogous contributions in other cases
considered earlier.
The result is
\beq
\chi^\e_{1,\bc;a_m}(t^*) =  \frac{1}{\sqrt{2\pi} (F^\e_1)^2} \frac{\cos(t^*-\pi/4)}{(t^*)^{3/2}} e^{-\gamma t^*}.
\label{ss_1}
\eeq
The second contribution, $\chi^\e_{1,\bc;b_m}(t^*)$,
comes from the vicinity of the point on the upper edge of the branch cut,
$s= x_{\mzs} - i (\gamma - 0+)$,
where there would be a ZS pole in the absence of damping. The real $x_{\mzs}$ is the solution of
\beq
\frac{1 + F^\e_1}{F^\e_1} = -2x^2_{\mzs} + 2 \frac{x^3_{\mzs}}{\sqrt{x^2_{\mzs} -1}}.
\label{ss_2}
\eeq
At $F^\e_1 =3/5$,  $x_{\mzs} = 2/\sqrt{3}$. For larger $F^\e_1$, $x_{\mzs}$
increases monotonically
with
$F^\e_1$.  For $F^\e_1 \gg 1$,  $x_{\mzs} \approx (3 F^\e_1/4)^{1/2}$.
For $s$ near  $x_{\mzs} - i (\gamma - 0+)$,
\beq
\chi^\e_1 (s) \approx -\frac{Q_1 (x_{\mzs})}{(F^\e_1)^2} \frac{1}{s -x_{\mzs} + i \gamma Q_2 (x_{\mzs})},
\label{ss_3}
\eeq
where
\bea
&& Q_1 (x_{\mzs}) = \frac{(x^2_{\mzs}-1)^{3/2}}{4 x_{\mzs}(x^2_{\mzs}-1)^{3/2} - 2 x^2_{\mzs} (2 x^2_{\mzs} -3)} \nonumber \\
&& Q_2(x_{\mzs}) = \frac{x^2_{\mzs} (x_{\mzs} - \sqrt{x^2_{\mzs}-1})}{2(x^2_{\mzs}-1)^{3/2} -  x_{\mzs} (2 x^2_{\mzs} -3)}
\label{ss_4}
\eea
Eq. (\ref{ss_3}) is valid only for $s$ above the branch cut, i.e., for
$|\text{Im}s| < \gamma$. This is satisfied on the upper branch of the cut, but not on the lower branch.

The function $Q_2 (x_{\mzs})$ satisfies $Q_2 (2/\sqrt{3}) =1$  and increases with $x_{\mzs}$ for larger $x_{\mzs}$, which correspond to $F^\e_1 > 3/5$.  At large $F^\e_1$, $Q_2 (x_{\mzs}) \approx F^\e_1/2$.   The condition $Q_2 (x_{\mzs}) >1$ implies that there is no pole in (\ref{ss_3}) above the branch cut, where this expression is valid.
Evaluating the branch cut contribution along $s = x -i(\gamma - 0+)$, we find that the largest piece comes from $x \approx x_{\mzs}$ and yields
\beq
\chi^\e_{1,\bc;b_m}(t^*) =  \frac{Q_1 (x_{\mzs})}{(F^\e_1)^2}  \sin(x_{\mzs} t^*) e^{- \gamma (Q_2 (x_{\mzs}))}
\label{ss_5}
\eeq
Combining (\ref{ss_1}) and (\ref{ss_5}), we see that in the range where a ZS pole is a mirage one,  $\chi^\e_1 (t^*)= - (\chi^\e_{1,\bc;a_m}(t^*) + \chi^\e_{1,\bc;b_m}(t^*))$ has a contribution oscillating with (dimensionless)
frequency $x_\mzs$ and the contribution oscillating with (dimensionless)
frequency equal to one.
When $F^\e_1 = O(1)$, the second contribution is the dominant one in some range of $t^* >1$,
because the first contribution contains $1/(t^*)^{3/2}$. However, above a certain $t^*$ the contribution from the branch point becomes the dominant one as it contains the smaller factor in the exponent.  This crossover
from oscillations with frequency $x_{\mzs}$ to oscillations with frequency $1$
provides a way to detect
a mirage pole experimentally.

For
$0 < F^\e_1 <3/5$, the ZS pole is
located
in the lower half-plane of frequency on the physical Rieman sheet. In this situation, $\chi^\e_1 (t^*)$ contains contributions
both
from the pole and from the branch cut. The combined contribution from the pole and the upper edge of the branch cut is
\beq
\chi^\e_{1}(t^*) =  2\frac{Q_1 (x_{\mzs})}{(F^\e_1)^2}  \sin(x_{\mzs} t^*) e^{- \gamma (Q_2 (x_{\mzs}))}
\label{ss_6}
\eeq
where now $0< x_{\mzs} < 2/\sqrt{3}$ and $Q_2 (x_{\mzs}) <1$.  The contribution from the branch points is
still given by (\ref{ss_1}). There is no crossover in this case because the exponential factor in the pole contribution is smaller than in the branch cut contribution. We note in passing that there is also a sign change between $\chi^\e_{1}(t^*)$ and $-\chi^\e_{1,\bc;b_m}(t^*)$ in (\ref{ss_5}), i.e., the phase of $\sin (x_{\mzs}) t^*)$ oscillations changes by $\pi$ between the regions where a ZS pole is a
conventional one and where it is a mirage one.

The same results can be obtained using the contour in Fig. \ref{fig:cont2}. For the contour of Fig. \ref{fig:cont2}, the pole contribution is non-zero and is given by
\begin{align}
  \label{eq:chi-pole-l-1-mirage}
  \chi_\pole(t^*) &= \frac{\sqrt{1-(s_1+i\gamma)^2}s_1^3}{F_1^\e{\displaystyle\prod_{j=2..4}(s_1-s_j)}}e^{-i s_1 t} + \cc, \\
\end{align}
where $s_1 = s_\mzs - i\gamma_\mzs$ is the mirage pole according to our conventions. This is just $-1$ times the result for a conventional ZS mode residing above the branch cut on the physical sheet, Eq. \eqref{eq:chi-pole-l-1}. The phase shift is due to the pole being on the unphysical sheet. The contribution of $\chi_{1,\bc}^\e(t^*)$ is dominated by the branch points and is given by
\beq
\chi^\e_{1,\bc}(t^*) =  \frac{1}{\sqrt{2\pi} (F^\e_1)^2} \frac{\cos(t^*-\pi/4)}{(t^*)^{3/2}} e^{-\gamma t^*},
\label{ss_1-2}
\eeq
The crossover time is
\begin{equation}
  t_{\crs,4} \sim \frac{1}{\gamma_{\mzs} - \gamma} \log\frac{1}{s_{\mzs}(\gamma_{\mzs} -\gamma)^{3/2}},
  \label{bb_12-m}
\end{equation}
i.e. it is analogous to the crossover time for a conventional pole with $\gamma_zs < \gamma$, see Eq. (\ref{bb_12}).

\subsection{Arbitrary $l$}
Our results for $l=0$ and $l=1$ can be readily generalized to any channel. Using the contour of Fig. \ref{fig:cont2}, we see that for a given channel with $2n$ poles on the Riemann surface, the solution is given by the contributions of mirage and conventional poles with $\gamma_\zs < \gamma$, along with the branch points contribution
\begin{equation}
  \label{eq:general-l}
  \chi_\bc(t^*) = Q_0 \sum_{j=1..2n} A_j e^{i\pi/4}Z(\sigma_j t^*),
\end{equation}
where $Z(a)$ is  given by Eq. \eqref{bb_5}, $A_j=\sum_{i\neq j}(\sigma_i-\sigma_j)^{-1}$ and $Q_0$ is a constant,
calculated directly from $\Delta\chi_l^\e(x)$
and
given by
\begin{equation}
  \label{eq:Q0-def}
  Q_0 = \lim_{x\to 0} \frac{\Delta\chi_l^\e(1-x)\prod_j(x+\sigma_j)}{\sqrt{2x}}.
\end{equation}
To study a crossover regime where a pole $s_1$ emerges from behind a branch cut, simply replace $e^{i\pi/4} Z(\sigma_j t^*)$ in \eqref{eq:general-l} by $\mathcal{Z}$, given in Eq. \eqref{eq:chi-bcut-l-1-2}.

\subsection{The case
  of comparable
  $F_0^\e$ and $F_1^\e$}
\label{sec:case-when-f_0e}

In the main text and in the previous sections, we assumed that one Landau parameter dominates over all others. In this section, we discuss what happens when two Landau parameters are comparable. We focus on the most physically relevant case when $F_0^\e$ and $F_1^\e$ dominate over all others, as can be expected for a generic interaction which decreases monotonically with momentum transfer. Our results can be readily generalized for the case of more nonzero $F_l^\e$'s.

When both $F_0^\e,F_1^\e$ are nonzero, the expression for the quasiparticle susceptiblity becomes more complex, since there are now cross terms in the ladder series.
Resumming the series, we obtain \cite{Wu2018,Zyuzin2018,Chubukov2019}
\begin{subequations}
  \begin{align}
    \chi^\e_0(s) & = \nu_F
                   \frac{\chi_0(1  +  F^\e_1\chi_1)  -
                   2 F^\e_1 \chi^2_{01}}
                   {(1 + F^\e_0 \chi_0)(1  +  F^\e_1\chi_1) -
                   2 F^\e_0 F^\e_1 \chi^2_{01}
                   },
                   \label{eq:chi0F0F1}\\
    \chi^\e_1(s) & = \nu_F
                    \frac{\chi_1(1  +  F^\e_0\chi_0)  -
                   2 F^\e_0 \chi^2_{01}}
                   {(1 + F^\e_0 \chi_0)(1  +  F^\e_1\chi_1) -
                   2 F^\e_0 F^\e_1 \chi^2_{01}
                   },
                  \label{eq:chi1F0F1}
  \end{align}
\end{subequations}
where $\chi_0$ and $\chi_1$ are given by Eqs.~\eqref{eq:chi-0} and \eqref{eq:chi-1-free}, respectively, while $\chi_{01}(s)$ is the fermion bubble with $l=0$ and $l=1$ form-factors at the vertices
\begin{equation}
  \label{eq:chi-01-free}
  \chi_{01}(s) = s\frac{1 + i \frac{s}{\sqrt{1-(s+i\gamma)^2}}}{1-\frac{\gamma}{\sqrt{1-(s+i\gamma)^2}}}.
\end{equation}
The equations for the poles in the $l=0$ and (longitudinal) $l=1$ channels are the same because  Eqs.~\eqref{eq:chi0F0F1}
and \eqref{eq:chi1F0F1}  have the same denominator.
 (The pole in the transverse $l=1$ channel is different.) The solution of
 \bea
 (1 + F^\e_0 \chi_0)(1  +  F^\e_1\chi_1) = 2 F^\e_0 F^\e_1 \chi^2_{01}
 \eea
 interpolates smoothly between the limits of  $|F^\e_0|\gg |F^\e_1|$ and $|F^\e_0|\ll |F^\e_1|$, studied in the previous sections.
As a result,
the behavior of the poles for the case of comparable $F_0^\e$ and $F^\e_1$ does not change qualitatively.  A new element, however, is that the mirage mode occurs both in the $l=0$ and $l=1$ channels (again, because they have a common pole). Also, the conditions for the existence of the mirage mode become less stringent compared to the  $F_0^\e=0$ case, when the mirage mode occurs only in the $l=1$ channel  and for $F_1^\e>3/5$. If $F_0^\e\neq 0$, the mirage mode occurs already for smaller values of $F_1^\e$, e.g., for $F_1^\e>0.15$ if $F_1^\e=1$.

For a charged FL, the situation is somewhat different. The new diagrammatic element are the chains of bubbles connected by the unscreened Coulomb interaction, $U_q=2\pi e^2/q$.  Such chains
 are present
 in the $l=0$ charge channel and in the $l\geq 1$ longitudinal charge channel, but not in the transverse charge channel and the spin channel. Each bubble in the chain is
  renormalized by a FL interaction, parameterized by the Landau function. The Landau function comprises infinite series of diagrams containing the {\em screened} Coulomb interaction.  Resumming the diagrammatic series, one obtains the full charge susceptibilities
   in the form
\begin{subequations}
\bea
\tilde\chi^c_0(q,\omega)&=&\frac{\chi_0^c(q,\omega)}{1-U_q\chi_0^c(q,\omega)}\label{Ca},\\
\tilde\chi^c_l(q,\omega)&=&\chi^c_l(q,\omega)+\frac{(\chi_{l0}^c(q,\omega))^2}{1-U_q\chi_0^c(q,\omega)},\,l\geq 1,\label{Cb}
\eea
\end{subequations}
where $\chi^c_l(q,\omega)$ is the quasiparticle susceptibility renormalized by the FL interaction and $\chi^c_{l0}(q,\omega)$ is the
``mixed"
 quasiparticle susceptibility with vertices at the opposite corners given by $\sqrt{2}\cos l\theta$ and $1$, correspondingly. The pole of (\ref{Ca}) is a 2D, $\sqrt{q}$ plasmon,
  whose group velocity is renormalized by the FL interaction \cite{Levitov2013}.
   This is the only collective mode in the $l=0$ charge channel.
   In the channels with
   $l\geq 1$
    there are two kinds of collective modes: the acoustic ZS modes, which correspond to the pole of the first term in Eq.~(\ref{Cb}), and the plasmon mode, which correspond to the pole of the second term in this equation. Note that the longitudinal ZS modes
    exist
     for any repulsive FL interaction, as opposed to the case of transverse ZS modes, which
     occur
       only if the FL interaction exceeds
      certain threshold \cite{Sodemann2019}.

\bibliography{FLtheory_fake}

\begin{thebibliography}{43}%
\makeatletter
\providecommand \@ifxundefined [1]{%
 \@ifx{#1\undefined}
}%
\providecommand \@ifnum [1]{%
 \ifnum #1\expandafter \@firstoftwo
 \else \expandafter \@secondoftwo
 \fi
}%
\providecommand \@ifx [1]{%
 \ifx #1\expandafter \@firstoftwo
 \else \expandafter \@secondoftwo
 \fi
}%
\providecommand \natexlab [1]{#1}%
\providecommand \enquote  [1]{``#1''}%
\providecommand \bibnamefont  [1]{#1}%
\providecommand \bibfnamefont [1]{#1}%
\providecommand \citenamefont [1]{#1}%
\providecommand \href@noop [0]{\@secondoftwo}%
\providecommand \href [0]{\begingroup \@sanitize@url \@href}%
\providecommand \@href[1]{\@@startlink{#1}\@@href}%
\providecommand \@@href[1]{\endgroup#1\@@endlink}%
\providecommand \@sanitize@url [0]{\catcode `\\12\catcode `\$12\catcode
  `\&12\catcode `\#12\catcode `\^12\catcode `\_12\catcode `\%12\relax}%
\providecommand \@@startlink[1]{}%
\providecommand \@@endlink[0]{}%
\providecommand \url  [0]{\begingroup\@sanitize@url \@url }%
\providecommand \@url [1]{\endgroup\@href {#1}{\urlprefix }}%
\providecommand \urlprefix  [0]{URL }%
\providecommand \Eprint [0]{\href }%
\providecommand \doibase [0]{http://dx.doi.org/}%
\providecommand \selectlanguage [0]{\@gobble}%
\providecommand \bibinfo  [0]{\@secondoftwo}%
\providecommand \bibfield  [0]{\@secondoftwo}%
\providecommand \translation [1]{[#1]}%
\providecommand \BibitemOpen [0]{}%
\providecommand \bibitemStop [0]{}%
\providecommand \bibitemNoStop [0]{.\EOS\space}%
\providecommand \EOS [0]{\spacefactor3000\relax}%
\providecommand \BibitemShut  [1]{\csname bibitem#1\endcsname}%
\let\auto@bib@innerbib\@empty
\bibitem [{\citenamefont {{A. A. Abrikosov, L. P. Gorkov, and I. E.
  Dzyaloshinski}}(1975)}]{Abrikosov1975}%
  \BibitemOpen
  \bibfield  {author} {\bibinfo {author} {\bibnamefont {{A. A. Abrikosov, L. P.
  Gorkov, and I. E. Dzyaloshinski}}},\ }\href@noop {} {\emph {\bibinfo {title}
  {Methods of Quantum Field Theory in Statistical Physics}}},\ Dover Books on
  Physics Series\ (\bibinfo  {publisher} {Dover Publications},\ \bibinfo {year}
  {1975})\BibitemShut {NoStop}%
\bibitem [{\citenamefont {{E. M. Lifshitz and L. P.
  Pitaevskii}}(1980)}]{Landau1980}%
  \BibitemOpen
  \bibfield  {author} {\bibinfo {author} {\bibnamefont {{E. M. Lifshitz and L.
  P. Pitaevskii}}},\ }\href@noop {} {\emph {\bibinfo {title} {Landau and
  Lifshitz Course of Theoretical Physics, v. IX: Statistical Physics, Part
  2}}},\ \bibinfo {number} {v. 9}\ (\bibinfo {year} {1980})\BibitemShut
  {NoStop}%
\bibitem [{\citenamefont {Baym}\ and\ \citenamefont
  {Pethick}(1991)}]{Baym1991}%
  \BibitemOpen
  \bibfield  {author} {\bibinfo {author} {\bibfnamefont {G.}~\bibnamefont
  {Baym}}\ and\ \bibinfo {author} {\bibfnamefont {C.~J.}\ \bibnamefont
  {Pethick}},\ }\href@noop {} {\emph {\bibinfo {title} {{Landau Fermi-Liquid
  Theory: Concepts and Applications}}}}\ (\bibinfo  {publisher} {John Wiley and
  Sons},\ \bibinfo {year} {1991})\BibitemShut {NoStop}%
\bibitem [{\citenamefont {Nozi{\`e}res}\ and\ \citenamefont
  {Pines}(1999)}]{Nozieres1999}%
  \BibitemOpen
  \bibfield  {author} {\bibinfo {author} {\bibfnamefont {P.}~\bibnamefont
  {Nozi{\`e}res}}\ and\ \bibinfo {author} {\bibfnamefont {D.}~\bibnamefont
  {Pines}},\ }\href@noop {} {\emph {\bibinfo {title} {{Theory of Quantum
  Liquids}}}}\ (\bibinfo  {publisher} {Hachette UK},\ \bibinfo {year}
  {1999})\BibitemShut {NoStop}%
\bibitem [{\citenamefont {Abel}\ \emph {et~al.}(1966)\citenamefont {Abel},
  \citenamefont {Anderson},\ and\ \citenamefont {Wheatley}}]{Abel1966}%
  \BibitemOpen
  \bibfield  {author} {\bibinfo {author} {\bibfnamefont {W.~R.}\ \bibnamefont
  {Abel}}, \bibinfo {author} {\bibfnamefont {A.~C.}\ \bibnamefont {Anderson}},
  \ and\ \bibinfo {author} {\bibfnamefont {J.~C.}\ \bibnamefont {Wheatley}},\
  }\href {\doibase 10.1103/PhysRevLett.17.74} {\bibfield  {journal} {\bibinfo
  {journal} {Phys. Rev. Lett.}\ }\textbf {\bibinfo {volume} {17}},\ \bibinfo
  {pages} {74} (\bibinfo {year} {1966})}\BibitemShut {NoStop}%
\bibitem [{\citenamefont {Pethick}\ and\ \citenamefont
  {Ravenhall}(1988)}]{Pethick1988}%
  \BibitemOpen
  \bibfield  {author} {\bibinfo {author} {\bibfnamefont {C.}~\bibnamefont
  {Pethick}}\ and\ \bibinfo {author} {\bibfnamefont {D.}~\bibnamefont
  {Ravenhall}},\ }\href@noop {} {\bibfield  {journal} {\bibinfo  {journal}
  {Annals of Physics}\ }\textbf {\bibinfo {volume} {183}},\ \bibinfo {pages}
  {131} (\bibinfo {year} {1988})}\BibitemShut {NoStop}%
\bibitem [{\citenamefont {Pomeranchuk}(1959)}]{Pomeranchuk1959}%
  \BibitemOpen
  \bibfield  {author} {\bibinfo {author} {\bibfnamefont {I.}~\bibnamefont
  {Pomeranchuk}},\ }\href@noop {} {\bibfield  {journal} {\bibinfo  {journal}
  {Sov. Phys. JETP}\ }\textbf {\bibinfo {volume} {8}},\ \bibinfo {pages} {361}
  (\bibinfo {year} {1959})}\BibitemShut {NoStop}%
\bibitem [{\citenamefont {Oganesyan}\ \emph {et~al.}(2001)\citenamefont
  {Oganesyan}, \citenamefont {Kivelson},\ and\ \citenamefont
  {Fradkin}}]{Oganesyan2001}%
  \BibitemOpen
  \bibfield  {author} {\bibinfo {author} {\bibfnamefont {V.}~\bibnamefont
  {Oganesyan}}, \bibinfo {author} {\bibfnamefont {S.~A.}\ \bibnamefont
  {Kivelson}}, \ and\ \bibinfo {author} {\bibfnamefont {E.}~\bibnamefont
  {Fradkin}},\ }\href {\doibase 10.1103/PhysRevB.64.195109} {\bibfield
  {journal} {\bibinfo  {journal} {Phys. Rev. B}\ }\textbf {\bibinfo {volume}
  {64}},\ \bibinfo {pages} {195109} (\bibinfo {year} {2001})}\BibitemShut
  {NoStop}%
\bibitem [{\citenamefont {Dell'Anna}\ and\ \citenamefont
  {Metzner}(2006)}]{DellAnna2006}%
  \BibitemOpen
  \bibfield  {author} {\bibinfo {author} {\bibfnamefont {L.}~\bibnamefont
  {Dell'Anna}}\ and\ \bibinfo {author} {\bibfnamefont {W.}~\bibnamefont
  {Metzner}},\ }\href {\doibase 10.1103/PhysRevB.73.045127} {\bibfield
  {journal} {\bibinfo  {journal} {Phys. Rev. B}\ }\textbf {\bibinfo {volume}
  {73}},\ \bibinfo {pages} {045127} (\bibinfo {year} {2006})}\BibitemShut
  {NoStop}%
\bibitem [{\citenamefont {Maslov}\ and\ \citenamefont
  {Chubukov}(2010)}]{Maslov2010}%
  \BibitemOpen
  \bibfield  {author} {\bibinfo {author} {\bibfnamefont {D.~L.}\ \bibnamefont
  {Maslov}}\ and\ \bibinfo {author} {\bibfnamefont {A.~V.}\ \bibnamefont
  {Chubukov}},\ }\href {\doibase 10.1103/PhysRevB.81.045110} {\bibfield
  {journal} {\bibinfo  {journal} {Phys. Rev. B}\ }\textbf {\bibinfo {volume}
  {81}},\ \bibinfo {pages} {045110} (\bibinfo {year} {2010})}\BibitemShut
  {NoStop}%
\bibitem [{\citenamefont {Watanabe}\ and\ \citenamefont
  {Vishwanath}(2014)}]{Watanabe2014}%
  \BibitemOpen
  \bibfield  {author} {\bibinfo {author} {\bibfnamefont {H.}~\bibnamefont
  {Watanabe}}\ and\ \bibinfo {author} {\bibfnamefont {A.}~\bibnamefont
  {Vishwanath}},\ }\href {\doibase 10.1073/pnas.1415592111} {\bibfield
  {journal} {\bibinfo  {journal} {Proc. Natl. Acad. Sci. USA}\ }\textbf
  {\bibinfo {volume} {111}},\ \bibinfo {pages} {16314} (\bibinfo {year}
  {2014})}\BibitemShut {NoStop}%
\bibitem [{\citenamefont {Kiselev}\ \emph {et~al.}(2017)\citenamefont
  {Kiselev}, \citenamefont {Scheurer}, \citenamefont {W\"olfle},\ and\
  \citenamefont {Schmalian}}]{Kiselev2017}%
  \BibitemOpen
  \bibfield  {author} {\bibinfo {author} {\bibfnamefont {E.~I.}\ \bibnamefont
  {Kiselev}}, \bibinfo {author} {\bibfnamefont {M.~S.}\ \bibnamefont
  {Scheurer}}, \bibinfo {author} {\bibfnamefont {P.}~\bibnamefont {W\"olfle}},
  \ and\ \bibinfo {author} {\bibfnamefont {J.}~\bibnamefont {Schmalian}},\
  }\href {\doibase 10.1103/PhysRevB.95.125122} {\bibfield  {journal} {\bibinfo
  {journal} {Phys. Rev. B}\ }\textbf {\bibinfo {volume} {95}},\ \bibinfo
  {pages} {125122} (\bibinfo {year} {2017})}\BibitemShut {NoStop}%
\bibitem [{\citenamefont {Chubukov}\ \emph {et~al.}(2018)\citenamefont
  {Chubukov}, \citenamefont {Klein},\ and\ \citenamefont
  {Maslov}}]{Chubukov2018}%
  \BibitemOpen
  \bibfield  {author} {\bibinfo {author} {\bibfnamefont {A.~V.}\ \bibnamefont
  {Chubukov}}, \bibinfo {author} {\bibfnamefont {A.}~\bibnamefont {Klein}}, \
  and\ \bibinfo {author} {\bibfnamefont {D.~L.}\ \bibnamefont {Maslov}},\
  }\href
  {https://urldefense.proofpoint.com/v2/url?u=https-3A__link.springer.com_article_10.1134_S1063776118110122&d=DwIGaQ&c=sJ6xIWYx-zLMB3EPkvcnVg&r=0buul-a0We0TvRMkSL-dD4cDqIj7U_RYTXT9igT7Q58&m=3-O3hXxIlpUU25SbkVQ7x-s9HCLdF_vRC1LqW_h_ozA&s=j82wHJdAdICTGshvkSI7E6JmNfmbQf2qFjQ_lpmhzxg&e=}
  {\bibfield  {journal} {\bibinfo  {journal} {JETP}\ }\textbf {\bibinfo
  {volume} {127}},\ \bibinfo {pages} {826} (\bibinfo {year}
  {2018})}\BibitemShut {NoStop}%
\bibitem [{\citenamefont {Klein}\ \emph {et~al.}(2019)\citenamefont {Klein},
  \citenamefont {Maslov}, \citenamefont {Pitaevskii},\ and\ \citenamefont
  {Chubukov}}]{Chubukov2019}%
  \BibitemOpen
  \bibfield  {author} {\bibinfo {author} {\bibfnamefont {A.}~\bibnamefont
  {Klein}}, \bibinfo {author} {\bibfnamefont {D.}~\bibnamefont {Maslov}},
  \bibinfo {author} {\bibfnamefont {L.}~\bibnamefont {Pitaevskii}}, \ and\
  \bibinfo {author} {\bibfnamefont {A.~V.}\ \bibnamefont {Chubukov}},\ }\href
  {https://urldefense.proofpoint.com/v2/url?u=https-3A__arxiv.org_abs_1908.04800&d=DwIGaQ&c=sJ6xIWYx-zLMB3EPkvcnVg&r=0buul-a0We0TvRMkSL-dD4cDqIj7U_RYTXT9igT7Q58&m=3-O3hXxIlpUU25SbkVQ7x-s9HCLdF_vRC1LqW_h_ozA&s=LBCvMjJcUYNnW0twBgIYfmwxJrYSOhlTFZL0Jz0aikk&e=}
  {\bibfield  {journal} {\bibinfo  {journal} {ArXiv:1908.04800}\ } (\bibinfo
  {year} {2019})}\BibitemShut {NoStop}%
\bibitem [{\citenamefont {Khoo}\ and\ \citenamefont
  {Villadiego}(2019)}]{Sodemann2019}%
  \BibitemOpen
  \bibfield  {author} {\bibinfo {author} {\bibfnamefont {J.~Y.}\ \bibnamefont
  {Khoo}}\ and\ \bibinfo {author} {\bibfnamefont {I.~S.}\ \bibnamefont
  {Villadiego}},\ }\href {\doibase 10.1103/PhysRevB.99.075434} {\bibfield
  {journal} {\bibinfo  {journal} {Phys. Rev. B}\ }\textbf {\bibinfo {volume}
  {99}},\ \bibinfo {pages} {075434} (\bibinfo {year} {2019})}\BibitemShut
  {NoStop}%
\bibitem [{SM()}]{SM}%
  \BibitemOpen
  \href@noop {} {}\bibinfo {note} {{See Supplementary Material for
  details}}\BibitemShut {NoStop}%
\bibitem [{\citenamefont {Nehari}(1952)}]{Nehari1952}%
  \BibitemOpen
  \bibfield  {author} {\bibinfo {author} {\bibfnamefont {Z.}~\bibnamefont
  {Nehari}},\ }\href@noop {} {\emph {\bibinfo {title} {Conformal mapping}}}\
  (\bibinfo  {publisher} {New York : McGraw-Hill},\ \bibinfo {year}
  {1952})\BibitemShut {NoStop}%
\bibitem [{\citenamefont {Farkas}\ and\ \citenamefont
  {Kra}(1992)}]{Farkas1992}%
  \BibitemOpen
  \bibfield  {author} {\bibinfo {author} {\bibfnamefont {H.~M.}\ \bibnamefont
  {Farkas}}\ and\ \bibinfo {author} {\bibfnamefont {I.}~\bibnamefont {Kra}},\
  }\href@noop {} {\emph {\bibinfo {title} {Riemann surfaces (2nd ed., Graduate
  texts in mathematics ; 71).}}}\ (\bibinfo  {publisher} {New York:
  Springer-Verlag.},\ \bibinfo {year} {1992})\BibitemShut {NoStop}%
\bibitem [{\citenamefont {Weyl}(1964)}]{Weyl1964}%
  \BibitemOpen
  \bibfield  {author} {\bibinfo {author} {\bibfnamefont {H.}~\bibnamefont
  {Weyl}},\ }\href@noop {} {\emph {\bibinfo {title} {The concept of a Riemann
  surface (3d ed., 1955.. ed., ADIWES international series in mathematics).}}}\
  (\bibinfo  {publisher} {Reading, Mass.: Addison-Wesley Pub.},\ \bibinfo
  {year} {1964})\BibitemShut {NoStop}%
\bibitem [{\citenamefont {Fal'ko}\ and\ \citenamefont
  {Khmel'nitskii}(1989)}]{Falko1989}%
  \BibitemOpen
  \bibfield  {author} {\bibinfo {author} {\bibfnamefont {V.~I.}\ \bibnamefont
  {Fal'ko}}\ and\ \bibinfo {author} {\bibfnamefont {D.~E.}\ \bibnamefont
  {Khmel'nitskii}},\ }\href
  {https://urldefense.proofpoint.com/v2/url?u=https-3A__ui.adsabs.harvard.edu_-255C-23abs_1989ZhETF..95.1988F&d=DwIGaQ&c=sJ6xIWYx-zLMB3EPkvcnVg&r=0buul-a0We0TvRMkSL-dD4cDqIj7U_RYTXT9igT7Q58&m=csoCZhHjjfp5hpbLJarQwXPxH6HdLk9MvNYzFEPKXNg&s=-BY4RwdwsvDoADm9_M_lq_kufngY1g9dtYH7ShM2f9w&e=}
  {\bibfield  {journal} {\bibinfo  {journal} {JETP}\ }\textbf {\bibinfo
  {volume} {68}},\ \bibinfo {pages} {1150} (\bibinfo {year}
  {1989})}\BibitemShut {NoStop}%
\bibitem [{\citenamefont {Oriekhov}\ and\ \citenamefont
  {Levitov}(2019)}]{Oriekhov2019}%
  \BibitemOpen
  \bibfield  {author} {\bibinfo {author} {\bibfnamefont {D.~O.}\ \bibnamefont
  {Oriekhov}}\ and\ \bibinfo {author} {\bibfnamefont {L.~S.}\ \bibnamefont
  {Levitov}},\ }\href
  {https://urldefense.proofpoint.com/v2/url?u=https-3A__ui.adsabs.harvard.edu_abs_2019arXiv190310648O&d=DwIGaQ&c=sJ6xIWYx-zLMB3EPkvcnVg&r=0buul-a0We0TvRMkSL-dD4cDqIj7U_RYTXT9igT7Q58&m=csoCZhHjjfp5hpbLJarQwXPxH6HdLk9MvNYzFEPKXNg&s=LsEOnZlB2C9d6gH5nG0pTwxoMgh9Q07Cf20vBPZ3heI&e=}
  {\bibfield  {journal} {\bibinfo  {journal} {arXiv:1903.10648}\ } (\bibinfo
  {year} {2019})}\BibitemShut {NoStop}%
\bibitem [{\citenamefont {Krausz}\ and\ \citenamefont
  {Ivanov}(2009)}]{Krausz2009}%
  \BibitemOpen
  \bibfield  {author} {\bibinfo {author} {\bibfnamefont {F.}~\bibnamefont
  {Krausz}}\ and\ \bibinfo {author} {\bibfnamefont {M.}~\bibnamefont
  {Ivanov}},\ }\href {\doibase 10.1103/RevModPhys.81.163} {\bibfield  {journal}
  {\bibinfo  {journal} {Rev. Mod. Phys.}\ }\textbf {\bibinfo {volume} {81}},\
  \bibinfo {pages} {163} (\bibinfo {year} {2009})}\BibitemShut {NoStop}%
\bibitem [{\citenamefont {Mitrano}\ \emph {et~al.}(2016)\citenamefont
  {Mitrano}, \citenamefont {Cantaluppi}, \citenamefont {Nicoletti},
  \citenamefont {Kaiser}, \citenamefont {Perucchi}, \citenamefont {Lupi},
  \citenamefont {Di~Pietro}, \citenamefont {Pontiroli}, \citenamefont
  {Ricc{\`o}}, \citenamefont {Clark}, \citenamefont {Jaksch},\ and\
  \citenamefont {Cavalleri}}]{Mitrano2016}%
  \BibitemOpen
  \bibfield  {author} {\bibinfo {author} {\bibfnamefont {M.}~\bibnamefont
  {Mitrano}}, \bibinfo {author} {\bibfnamefont {A.}~\bibnamefont {Cantaluppi}},
  \bibinfo {author} {\bibfnamefont {D.}~\bibnamefont {Nicoletti}}, \bibinfo
  {author} {\bibfnamefont {S.}~\bibnamefont {Kaiser}}, \bibinfo {author}
  {\bibfnamefont {A.}~\bibnamefont {Perucchi}}, \bibinfo {author}
  {\bibfnamefont {S.}~\bibnamefont {Lupi}}, \bibinfo {author} {\bibfnamefont
  {P.}~\bibnamefont {Di~Pietro}}, \bibinfo {author} {\bibfnamefont
  {D.}~\bibnamefont {Pontiroli}}, \bibinfo {author} {\bibfnamefont
  {M.}~\bibnamefont {Ricc{\`o}}}, \bibinfo {author} {\bibfnamefont {S.~R.}\
  \bibnamefont {Clark}}, \bibinfo {author} {\bibfnamefont {D.}~\bibnamefont
  {Jaksch}}, \ and\ \bibinfo {author} {\bibfnamefont {A.}~\bibnamefont
  {Cavalleri}},\ }\href
  {https://urldefense.proofpoint.com/v2/url?u=https-3A__doi.org_10.1038_nature16522&d=DwIGaQ&c=sJ6xIWYx-zLMB3EPkvcnVg&r=0buul-a0We0TvRMkSL-dD4cDqIj7U_RYTXT9igT7Q58&m=3-O3hXxIlpUU25SbkVQ7x-s9HCLdF_vRC1LqW_h_ozA&s=nltOZn6Ec63Merb2-s5brr2xFFVYEFLJainr98x-p0g&e=}
  {\bibfield  {journal} {\bibinfo  {journal} {Nature}\ }\textbf {\bibinfo
  {volume} {530}},\ \bibinfo {pages} {461} (\bibinfo {year}
  {2016})}\BibitemShut {NoStop}%
\bibitem [{\citenamefont {Giannetti}\ \emph {et~al.}(2016)\citenamefont
  {Giannetti}, \citenamefont {Capone}, \citenamefont {Fausti}, \citenamefont
  {Fabrizio}, \citenamefont {Parmigiani},\ and\ \citenamefont
  {Mihailovic}}]{Giannetti2016}%
  \BibitemOpen
  \bibfield  {author} {\bibinfo {author} {\bibfnamefont {C.}~\bibnamefont
  {Giannetti}}, \bibinfo {author} {\bibfnamefont {M.}~\bibnamefont {Capone}},
  \bibinfo {author} {\bibfnamefont {D.}~\bibnamefont {Fausti}}, \bibinfo
  {author} {\bibfnamefont {M.}~\bibnamefont {Fabrizio}}, \bibinfo {author}
  {\bibfnamefont {F.}~\bibnamefont {Parmigiani}}, \ and\ \bibinfo {author}
  {\bibfnamefont {D.}~\bibnamefont {Mihailovic}},\ }\bibfield  {booktitle}
  {\emph {\bibinfo {booktitle} {Advances in Physics}},\ }\href {\doibase
  10.1080/00018732.2016.1194044} {\bibfield  {journal} {\bibinfo  {journal}
  {Advances in Physics}\ }\textbf {\bibinfo {volume} {65}},\ \bibinfo {pages}
  {58} (\bibinfo {year} {2016})}\BibitemShut {NoStop}%
\bibitem [{\citenamefont {Gandolfi}\ \emph {et~al.}(2017)\citenamefont
  {Gandolfi}, \citenamefont {Celardo}, \citenamefont {Borgonovi}, \citenamefont
  {Ferrini}, \citenamefont {Avella}, \citenamefont {Banfi},\ and\ \citenamefont
  {Giannetti}}]{Gandolfi2017}%
  \BibitemOpen
  \bibfield  {author} {\bibinfo {author} {\bibfnamefont {M.}~\bibnamefont
  {Gandolfi}}, \bibinfo {author} {\bibfnamefont {G.~L.}\ \bibnamefont
  {Celardo}}, \bibinfo {author} {\bibfnamefont {F.}~\bibnamefont {Borgonovi}},
  \bibinfo {author} {\bibfnamefont {G.}~\bibnamefont {Ferrini}}, \bibinfo
  {author} {\bibfnamefont {A.}~\bibnamefont {Avella}}, \bibinfo {author}
  {\bibfnamefont {F.}~\bibnamefont {Banfi}}, \ and\ \bibinfo {author}
  {\bibfnamefont {C.}~\bibnamefont {Giannetti}},\ }\href
  {https://urldefense.proofpoint.com/v2/url?u=http-3A__stacks.iop.org_1402-2D4896_92_i-3D3_a-3D034004&d=DwIGaQ&c=sJ6xIWYx-zLMB3EPkvcnVg&r=0buul-a0We0TvRMkSL-dD4cDqIj7U_RYTXT9igT7Q58&m=3-O3hXxIlpUU25SbkVQ7x-s9HCLdF_vRC1LqW_h_ozA&s=HXcGoL3-ymuhbnB6__h_HyBVj1wWiSqsamDpGhSL7G4&e=}
  {\bibfield  {journal} {\bibinfo  {journal} {Physica Scripta}\ }\textbf
  {\bibinfo {volume} {92}},\ \bibinfo {pages} {034004} (\bibinfo {year}
  {2017})}\BibitemShut {NoStop}%
\bibitem [{\citenamefont {Nosarzewski}\ \emph {et~al.}(2017)\citenamefont
  {Nosarzewski}, \citenamefont {Moritz}, \citenamefont {Freericks},
  \citenamefont {Kemper},\ and\ \citenamefont {Devereaux}}]{Nosarzewski2017}%
  \BibitemOpen
  \bibfield  {author} {\bibinfo {author} {\bibfnamefont {B.}~\bibnamefont
  {Nosarzewski}}, \bibinfo {author} {\bibfnamefont {B.}~\bibnamefont {Moritz}},
  \bibinfo {author} {\bibfnamefont {J.~K.}\ \bibnamefont {Freericks}}, \bibinfo
  {author} {\bibfnamefont {A.~F.}\ \bibnamefont {Kemper}}, \ and\ \bibinfo
  {author} {\bibfnamefont {T.~P.}\ \bibnamefont {Devereaux}},\ }\href {\doibase
  10.1103/PhysRevB.96.184518} {\bibfield  {journal} {\bibinfo  {journal} {Phys.
  Rev. B}\ }\textbf {\bibinfo {volume} {96}},\ \bibinfo {pages} {184518}
  (\bibinfo {year} {2017})}\BibitemShut {NoStop}%
\bibitem [{\citenamefont {Nicoletti}\ and\ \citenamefont
  {Cavalleri}(2016)}]{Nicoletti2016}%
  \BibitemOpen
  \bibfield  {author} {\bibinfo {author} {\bibfnamefont {D.}~\bibnamefont
  {Nicoletti}}\ and\ \bibinfo {author} {\bibfnamefont {A.}~\bibnamefont
  {Cavalleri}},\ }\bibfield  {booktitle} {\emph {\bibinfo {booktitle} {Advances
  in Optics and Photonics}},\ }\href
  {https://urldefense.proofpoint.com/v2/url?u=http-3A__aop.osa.org_abstract.cfm-3FURI-3Daop-2D8-2D3-2D401&d=DwIGaQ&c=sJ6xIWYx-zLMB3EPkvcnVg&r=0buul-a0We0TvRMkSL-dD4cDqIj7U_RYTXT9igT7Q58&m=3-O3hXxIlpUU25SbkVQ7x-s9HCLdF_vRC1LqW_h_ozA&s=zFwVz-wmLbakWn7pxnVAiPSCMUU_rrmrSEFj19ru1l4&e=}
  {\bibfield  {journal} {\bibinfo  {journal} {Adv. Opt. Photon.}\ }\textbf
  {\bibinfo {volume} {8}},\ \bibinfo {pages} {401} (\bibinfo {year}
  {2016})}\BibitemShut {NoStop}%
\bibitem [{\citenamefont {von Hoegen}\ \emph {et~al.}(2018)\citenamefont {von
  Hoegen}, \citenamefont {Mankowsky}, \citenamefont {Fechner}, \citenamefont
  {F{\"o}rst},\ and\ \citenamefont {Cavalleri}}]{Hoegen2018}%
  \BibitemOpen
  \bibfield  {author} {\bibinfo {author} {\bibfnamefont {A.}~\bibnamefont {von
  Hoegen}}, \bibinfo {author} {\bibfnamefont {R.}~\bibnamefont {Mankowsky}},
  \bibinfo {author} {\bibfnamefont {M.}~\bibnamefont {Fechner}}, \bibinfo
  {author} {\bibfnamefont {M.}~\bibnamefont {F{\"o}rst}}, \ and\ \bibinfo
  {author} {\bibfnamefont {A.}~\bibnamefont {Cavalleri}},\ }\href
  {https://urldefense.proofpoint.com/v2/url?u=http-3A__dx.doi.org_10.1038_nature25484&d=DwIGaQ&c=sJ6xIWYx-zLMB3EPkvcnVg&r=0buul-a0We0TvRMkSL-dD4cDqIj7U_RYTXT9igT7Q58&m=3-O3hXxIlpUU25SbkVQ7x-s9HCLdF_vRC1LqW_h_ozA&s=C8jKt7qkkqN_qw4ZHY7ZywbbHHjUJ5N6Eglfp9TJp4k&e=}
  {\bibfield  {journal} {\bibinfo  {journal} {Nature}\ }\textbf {\bibinfo
  {volume} {555}},\ \bibinfo {pages} {79} (\bibinfo {year} {2018})}\BibitemShut
  {NoStop}%
\bibitem [{\citenamefont {Mitra}(2018)}]{Mitra2019}%
  \BibitemOpen
  \bibfield  {author} {\bibinfo {author} {\bibfnamefont {A.}~\bibnamefont
  {Mitra}},\ }\href {\doibase 10.1146/annurev-conmatphys-031016-025451}
  {\bibfield  {journal} {\bibinfo  {journal} {Annu. Rev. Condens. Matter
  Phys.}\ }\textbf {\bibinfo {volume} {9}},\ \bibinfo {pages} {245} (\bibinfo
  {year} {2018})}\BibitemShut {NoStop}%
\bibitem [{\citenamefont {Zong}\ \emph {et~al.}(2019)\citenamefont {Zong},
  \citenamefont {Dolgirev}, \citenamefont {Kogar}, \citenamefont
  {Erge\ifmmode~\mbox{\c{c}}\else \c{c}\fi{}en}, \citenamefont {Yilmaz},
  \citenamefont {Bie}, \citenamefont {Rohwer}, \citenamefont {Tung},
  \citenamefont {Straquadine}, \citenamefont {Wang}, \citenamefont {Yang},
  \citenamefont {Shen}, \citenamefont {Li}, \citenamefont {Yang}, \citenamefont
  {Park}, \citenamefont {Hoffmann}, \citenamefont {Ofori-Okai}, \citenamefont
  {Kozina}, \citenamefont {Wen}, \citenamefont {Wang}, \citenamefont {Fisher},
  \citenamefont {Jarillo-Herrero},\ and\ \citenamefont {Gedik}}]{Gedik2019}%
  \BibitemOpen
  \bibfield  {author} {\bibinfo {author} {\bibfnamefont {A.}~\bibnamefont
  {Zong}}, \bibinfo {author} {\bibfnamefont {P.~E.}\ \bibnamefont {Dolgirev}},
  \bibinfo {author} {\bibfnamefont {A.}~\bibnamefont {Kogar}}, \bibinfo
  {author} {\bibfnamefont {E.}~\bibnamefont {Erge\ifmmode~\mbox{\c{c}}\else
  \c{c}\fi{}en}}, \bibinfo {author} {\bibfnamefont {M.~B.}\ \bibnamefont
  {Yilmaz}}, \bibinfo {author} {\bibfnamefont {Y.-Q.}\ \bibnamefont {Bie}},
  \bibinfo {author} {\bibfnamefont {T.}~\bibnamefont {Rohwer}}, \bibinfo
  {author} {\bibfnamefont {I.-C.}\ \bibnamefont {Tung}}, \bibinfo {author}
  {\bibfnamefont {J.}~\bibnamefont {Straquadine}}, \bibinfo {author}
  {\bibfnamefont {X.}~\bibnamefont {Wang}}, \bibinfo {author} {\bibfnamefont
  {Y.}~\bibnamefont {Yang}}, \bibinfo {author} {\bibfnamefont {X.}~\bibnamefont
  {Shen}}, \bibinfo {author} {\bibfnamefont {R.}~\bibnamefont {Li}}, \bibinfo
  {author} {\bibfnamefont {J.}~\bibnamefont {Yang}}, \bibinfo {author}
  {\bibfnamefont {S.}~\bibnamefont {Park}}, \bibinfo {author} {\bibfnamefont
  {M.~C.}\ \bibnamefont {Hoffmann}}, \bibinfo {author} {\bibfnamefont {B.~K.}\
  \bibnamefont {Ofori-Okai}}, \bibinfo {author} {\bibfnamefont {M.~E.}\
  \bibnamefont {Kozina}}, \bibinfo {author} {\bibfnamefont {H.}~\bibnamefont
  {Wen}}, \bibinfo {author} {\bibfnamefont {X.}~\bibnamefont {Wang}}, \bibinfo
  {author} {\bibfnamefont {I.~R.}\ \bibnamefont {Fisher}}, \bibinfo {author}
  {\bibfnamefont {P.}~\bibnamefont {Jarillo-Herrero}}, \ and\ \bibinfo {author}
  {\bibfnamefont {N.}~\bibnamefont {Gedik}},\ }\href {\doibase
  10.1103/PhysRevLett.123.097601} {\bibfield  {journal} {\bibinfo  {journal}
  {Phys. Rev. Lett.}\ }\textbf {\bibinfo {volume} {123}},\ \bibinfo {pages}
  {097601} (\bibinfo {year} {2019})}\BibitemShut {NoStop}%
\bibitem [{\citenamefont {Zala}\ \emph {et~al.}(2001)\citenamefont {Zala},
  \citenamefont {Narozhny},\ and\ \citenamefont {Aleiner}}]{Zala2001}%
  \BibitemOpen
  \bibfield  {author} {\bibinfo {author} {\bibfnamefont {G.}~\bibnamefont
  {Zala}}, \bibinfo {author} {\bibfnamefont {B.~N.}\ \bibnamefont {Narozhny}},
  \ and\ \bibinfo {author} {\bibfnamefont {I.~L.}\ \bibnamefont {Aleiner}},\
  }\href {\doibase 10.1103/PhysRevB.64.214204} {\bibfield  {journal} {\bibinfo
  {journal} {Phys. Rev. B}\ }\textbf {\bibinfo {volume} {64}},\ \bibinfo
  {pages} {214204} (\bibinfo {year} {2001})}\BibitemShut {NoStop}%
\bibitem [{foo()}]{footnote1}%
  \BibitemOpen
  \href@noop {} {}\bibinfo {note} {The $l=1$ transverse mode has recently been
  discussed in Refs. \onlinecite{Sodemann2019, Chubukov2019}.}\BibitemShut
  {Stop}%
\bibitem [{\citenamefont {Lucas}\ and\ \citenamefont
  {Das~Sarma}(2018)}]{Lucas2018}%
  \BibitemOpen
  \bibfield  {author} {\bibinfo {author} {\bibfnamefont {A.}~\bibnamefont
  {Lucas}}\ and\ \bibinfo {author} {\bibfnamefont {S.}~\bibnamefont
  {Das~Sarma}},\ }\href {\doibase 10.1103/PhysRevB.97.115449} {\bibfield
  {journal} {\bibinfo  {journal} {Phys. Rev. B}\ }\textbf {\bibinfo {volume}
  {97}},\ \bibinfo {pages} {115449} (\bibinfo {year} {2018})}\BibitemShut
  {NoStop}%
\bibitem [{\citenamefont {Mitrano}\ \emph {et~al.}(2019)\citenamefont
  {Mitrano}, \citenamefont {Lee}, \citenamefont {Husain}, \citenamefont
  {Delacretaz}, \citenamefont {Zhu}, \citenamefont {de~la Pena~Munoz},
  \citenamefont {Sun}, \citenamefont {Joe}, \citenamefont {Reid}, \citenamefont
  {Wandel}, \citenamefont {Coslovich}, \citenamefont {Schlotter}, \citenamefont
  {van Driel}, \citenamefont {Schneeloch}, \citenamefont {Gu}, \citenamefont
  {Hartnoll}, \citenamefont {Goldenfeld},\ and\ \citenamefont
  {Abbamonte}}]{Mitrano2019}%
  \BibitemOpen
  \bibfield  {author} {\bibinfo {author} {\bibfnamefont {M.}~\bibnamefont
  {Mitrano}}, \bibinfo {author} {\bibfnamefont {S.}~\bibnamefont {Lee}},
  \bibinfo {author} {\bibfnamefont {A.~A.}\ \bibnamefont {Husain}}, \bibinfo
  {author} {\bibfnamefont {L.}~\bibnamefont {Delacretaz}}, \bibinfo {author}
  {\bibfnamefont {M.}~\bibnamefont {Zhu}}, \bibinfo {author} {\bibfnamefont
  {G.}~\bibnamefont {de~la Pena~Munoz}}, \bibinfo {author} {\bibfnamefont
  {S.~X.-L.}\ \bibnamefont {Sun}}, \bibinfo {author} {\bibfnamefont {Y.~I.}\
  \bibnamefont {Joe}}, \bibinfo {author} {\bibfnamefont {A.~H.}\ \bibnamefont
  {Reid}}, \bibinfo {author} {\bibfnamefont {S.~F.}\ \bibnamefont {Wandel}},
  \bibinfo {author} {\bibfnamefont {G.}~\bibnamefont {Coslovich}}, \bibinfo
  {author} {\bibfnamefont {W.}~\bibnamefont {Schlotter}}, \bibinfo {author}
  {\bibfnamefont {T.}~\bibnamefont {van Driel}}, \bibinfo {author}
  {\bibfnamefont {J.}~\bibnamefont {Schneeloch}}, \bibinfo {author}
  {\bibfnamefont {G.~D.}\ \bibnamefont {Gu}}, \bibinfo {author} {\bibfnamefont
  {S.}~\bibnamefont {Hartnoll}}, \bibinfo {author} {\bibfnamefont
  {N.}~\bibnamefont {Goldenfeld}}, \ and\ \bibinfo {author} {\bibfnamefont
  {P.}~\bibnamefont {Abbamonte}},\ }\href
  {https://advances.sciencemag.org/content/5/8/eaax3346} {\bibfield  {journal}
  {\bibinfo  {journal} {Science Advances}\ }\textbf {\bibinfo {volume} {5}}
  (\bibinfo {year} {2019})}\BibitemShut {NoStop}%
\bibitem [{\citenamefont {Granroth}\ \emph {et~al.}(2018)\citenamefont
  {Granroth}, \citenamefont {An}, \citenamefont {Smith}, \citenamefont
  {Whitfield}, \citenamefont {Neuefeind}, \citenamefont {Lee}, \citenamefont
  {Zhou}, \citenamefont {Sedov}, \citenamefont {Peterson}, \citenamefont
  {Parizzi}, \citenamefont {Skorpenske}, \citenamefont {Hartman}, \citenamefont
  {Huq},\ and\ \citenamefont {Abernathy}}]{Granroth2018}%
  \BibitemOpen
  \bibfield  {author} {\bibinfo {author} {\bibfnamefont {G.~E.}\ \bibnamefont
  {Granroth}}, \bibinfo {author} {\bibfnamefont {K.}~\bibnamefont {An}},
  \bibinfo {author} {\bibfnamefont {H.~L.}\ \bibnamefont {Smith}}, \bibinfo
  {author} {\bibfnamefont {P.}~\bibnamefont {Whitfield}}, \bibinfo {author}
  {\bibfnamefont {J.~C.}\ \bibnamefont {Neuefeind}}, \bibinfo {author}
  {\bibfnamefont {J.}~\bibnamefont {Lee}}, \bibinfo {author} {\bibfnamefont
  {W.}~\bibnamefont {Zhou}}, \bibinfo {author} {\bibfnamefont {V.~N.}\
  \bibnamefont {Sedov}}, \bibinfo {author} {\bibfnamefont {P.~F.}\ \bibnamefont
  {Peterson}}, \bibinfo {author} {\bibfnamefont {A.}~\bibnamefont {Parizzi}},
  \bibinfo {author} {\bibfnamefont {H.}~\bibnamefont {Skorpenske}}, \bibinfo
  {author} {\bibfnamefont {S.~M.}\ \bibnamefont {Hartman}}, \bibinfo {author}
  {\bibfnamefont {A.}~\bibnamefont {Huq}}, \ and\ \bibinfo {author}
  {\bibfnamefont {D.~L.}\ \bibnamefont {Abernathy}},\ }\href {\doibase
  10.1107/S1600576718004727} {\bibfield  {journal} {\bibinfo  {journal} {J.
  Appl. Cryst.}\ }\textbf {\bibinfo {volume} {51}},\ \bibinfo {pages} {616}
  (\bibinfo {year} {2018})}\BibitemShut {NoStop}%
\bibitem [{\citenamefont {Tan}\ \emph {et~al.}(2005)\citenamefont {Tan},
  \citenamefont {Zhu}, \citenamefont {Stormer}, \citenamefont {Pfeiffer},
  \citenamefont {Baldwin},\ and\ \citenamefont {West}}]{Tan2005}%
  \BibitemOpen
  \bibfield  {author} {\bibinfo {author} {\bibfnamefont {Y.-W.}\ \bibnamefont
  {Tan}}, \bibinfo {author} {\bibfnamefont {J.}~\bibnamefont {Zhu}}, \bibinfo
  {author} {\bibfnamefont {H.~L.}\ \bibnamefont {Stormer}}, \bibinfo {author}
  {\bibfnamefont {L.~N.}\ \bibnamefont {Pfeiffer}}, \bibinfo {author}
  {\bibfnamefont {K.~W.}\ \bibnamefont {Baldwin}}, \ and\ \bibinfo {author}
  {\bibfnamefont {K.~W.}\ \bibnamefont {West}},\ }\href {\doibase
  10.1103/PhysRevLett.94.016405} {\bibfield  {journal} {\bibinfo  {journal}
  {Phys. Rev. Lett.}\ }\textbf {\bibinfo {volume} {94}},\ \bibinfo {pages}
  {016405} (\bibinfo {year} {2005})}\BibitemShut {NoStop}%
\bibitem [{\citenamefont {Tan}\ \emph {et~al.}(2006)\citenamefont {Tan},
  \citenamefont {Zhu}, \citenamefont {Stormer}, \citenamefont {Pfeiffer},
  \citenamefont {Baldwin},\ and\ \citenamefont {West}}]{Tan2006}%
  \BibitemOpen
  \bibfield  {author} {\bibinfo {author} {\bibfnamefont {Y.-W.}\ \bibnamefont
  {Tan}}, \bibinfo {author} {\bibfnamefont {J.}~\bibnamefont {Zhu}}, \bibinfo
  {author} {\bibfnamefont {H.~L.}\ \bibnamefont {Stormer}}, \bibinfo {author}
  {\bibfnamefont {L.~N.}\ \bibnamefont {Pfeiffer}}, \bibinfo {author}
  {\bibfnamefont {K.~W.}\ \bibnamefont {Baldwin}}, \ and\ \bibinfo {author}
  {\bibfnamefont {K.~W.}\ \bibnamefont {West}},\ }\href {\doibase
  10.1103/PhysRevB.73.045334} {\bibfield  {journal} {\bibinfo  {journal} {Phys.
  Rev. B}\ }\textbf {\bibinfo {volume} {73}},\ \bibinfo {pages} {045334}
  (\bibinfo {year} {2006})}\BibitemShut {NoStop}%
\bibitem [{\citenamefont {Leggett}(1965)}]{Leggett1965}%
  \BibitemOpen
  \bibfield  {author} {\bibinfo {author} {\bibfnamefont {A.~J.}\ \bibnamefont
  {Leggett}},\ }\href {\doibase 10.1103/PhysRev.140.A1869} {\bibfield
  {journal} {\bibinfo  {journal} {Phys. Rev.}\ }\textbf {\bibinfo {volume}
  {140}},\ \bibinfo {pages} {A1869} (\bibinfo {year} {1965})}\BibitemShut
  {NoStop}%
\bibitem [{\citenamefont {Wu}\ \emph {et~al.}(2018)\citenamefont {Wu},
  \citenamefont {Klein},\ and\ \citenamefont {Chubukov}}]{Wu2018}%
  \BibitemOpen
  \bibfield  {author} {\bibinfo {author} {\bibfnamefont {Y.-M.}\ \bibnamefont
  {Wu}}, \bibinfo {author} {\bibfnamefont {A.}~\bibnamefont {Klein}}, \ and\
  \bibinfo {author} {\bibfnamefont {A.~V.}\ \bibnamefont {Chubukov}},\ }\href
  {\doibase 10.1103/PhysRevB.97.165101} {\bibfield  {journal} {\bibinfo
  {journal} {Phys. Rev. B}\ }\textbf {\bibinfo {volume} {97}},\ \bibinfo
  {pages} {165101} (\bibinfo {year} {2018})}\BibitemShut {NoStop}%
\bibitem [{\citenamefont {Zyuzin}\ \emph {et~al.}(2018)\citenamefont {Zyuzin},
  \citenamefont {Sharma},\ and\ \citenamefont {Maslov}}]{Zyuzin2018}%
  \BibitemOpen
  \bibfield  {author} {\bibinfo {author} {\bibfnamefont {V.~A.}\ \bibnamefont
  {Zyuzin}}, \bibinfo {author} {\bibfnamefont {P.}~\bibnamefont {Sharma}}, \
  and\ \bibinfo {author} {\bibfnamefont {D.~L.}\ \bibnamefont {Maslov}},\
  }\href {\doibase 10.1103/PhysRevB.98.115139} {\bibfield  {journal} {\bibinfo
  {journal} {Phys. Rev. B}\ }\textbf {\bibinfo {volume} {98}},\ \bibinfo
  {pages} {115139} (\bibinfo {year} {2018})}\BibitemShut {NoStop}%
\bibitem [{Note1()}]{Note1}%
  \BibitemOpen
  \bibinfo {note} {We fit segments of the data at different $t^*$ onto a trial
  function $A \protect \qopname \relax o{cos}(t^* - \phi )/(t^*)^\alpha $,
  where $A,\phi ,\alpha $ are fitting parameters. We then fit $\phi
  (t^*/t_{\protect \text {cross}})$ to the prediction of Eq. \protect \textup
  {\hbox {\mathsurround \z@ \protect \normalfont (\ignorespaces \ref
  {eq:chi-branch-hidden-0-res}\unskip \@@italiccorr )}}.}\BibitemShut {Stop}%
\bibitem [{Note2()}]{Note2}%
  \BibitemOpen
  \bibinfo {note} {Note that since $s_{2,3}$ are not near the branch point at
  $1-i\gamma $, they have $\sigma _j \approx -2$ while the integral is
  dominated by the region $y \sim |\sigma _1|,|\sigma _4|$. However, their
  contribution is included in the complex conjugate term in $\chi _{1,{\protect
  \text {branch}}}$}\BibitemShut {NoStop}%
\bibitem [{\citenamefont {Levitov}\ \emph {et~al.}(2013)\citenamefont
  {Levitov}, \citenamefont {Shtyk},\ and\ \citenamefont
  {Feigelman}}]{Levitov2013}%
  \BibitemOpen
  \bibfield  {author} {\bibinfo {author} {\bibfnamefont {L.~S.}\ \bibnamefont
  {Levitov}}, \bibinfo {author} {\bibfnamefont {A.~V.}\ \bibnamefont {Shtyk}},
  \ and\ \bibinfo {author} {\bibfnamefont {M.~V.}\ \bibnamefont {Feigelman}},\
  }\href {\doibase 10.1103/PhysRevB.88.235403} {\bibfield  {journal} {\bibinfo
  {journal} {Phys. Rev. B}\ }\textbf {\bibinfo {volume} {88}},\ \bibinfo
  {pages} {235403} (\bibinfo {year} {2013})}\BibitemShut {NoStop}%
\end{thebibliography}%
\end{document}